# Multipurpose and Reusable Ultrathin Electronic Tattoos Based on PtSe$_2$ and PtTe$_2$

*Dmitry Kireev[1,2,\*], Emmanuel Okogbue[3,4], Jayanth RT[1], Tae-Jun Ko[3], Yeonwoong Jung[3,4,5], and Deji Akinwande[1,2,6]*

1 Department of Electrical and Computer Engineering, The University of Texas at Austin, Austin, Texas, 78758 USA

2 Microelectronics Research Center, The University of Texas, Austin, Texas, 78758 USA

3 NanoScience Technology Center, University of Central Florida, Orlando, Florida 32826, USA

4 Department of Electrical and Computer Engineering, University of Central Florida, Orlando, Florida 32816, USA

5 Department of Materials Science and Engineering, University of Central Florida, Orlando, Florida 32816, USA

6 Texas Materials Institute, The University of Texas at Austin, Austin, Texas, 78758 USA

\* Corresponding author: d.kireev@utexas.edu or kirdmitry@gmail.com

ABSTRACT    Wearable bioelectronics with emphasis on the research and development of advanced person-oriented biomedical devices have attracted immense interest in the last decade. Scientists and clinicians find it essential to utilize skin-worn smart tattoos for on-demand and ambulatory monitoring of an individual's vital signs. Here we report on the development of novel ultrathin platinum-based two-dimensional dichalcogenide (Pt-TMDs) based electronic tattoos as advanced building blocks of future wearable bioelectronics. We made these ultrathin electronic tattoos out of large-scale synthesized platinum diselenide ($PtSe_2$) and platinum ditelluride ($PtTe_2$) layered materials and used them for monitoring human physiological vital signs, such as the electrical activity of the heart and the brain, muscle contractions, eye movements, and temperature. We show that both materials can be used for these applications; yet, $PtTe_2$ was found to be the most suitable choice due to its metallic structure. In terms of sheet resistance, skin-contact, and electrochemical impedance, $PtTe_2$ outperforms state-of-the-art gold and graphene electronic tattoos and performs on par with medical-grade Ag/AgCl gel electrodes. The $PtTe_2$ tattoos show four times lower impedance and almost 100 times lower sheet resistance compared to monolayer graphene tattoos. One of the possible prompt implications of the work is perhaps in the development of advanced human-machine interfaces. To display the application, we built a multi-tattoo system that can easily distinguish eye movement and identify the direction of an individual's sight.

Atomically thin layered materials have recently emerged as an entirely novel class of materials with advanced, peculiar, and often unexpected properties.[1] Transition metal dichalcogenides (TMDs), with a generic $MX_2$ composition, are perhaps the most well studied of these two-dimensional materials, beyond graphene, up to date.[2] Changing the constituents of M and X in certain combinations is allowed where M is a transition metal (e.g., Mo, W, Pt, V, etc.), and X is a chalcogen (S, Se, Te). While most of the TMDs known to-date are semiconducting or metallic, they are found to possess a plurality of other properties for realizing fully 2D-based electronic devices with an entirely new set of properties. The layered structure associated with TMDs is of particular interest, and the electronic properties vary drastically from bulk to multi- and single-layered materials.[3,4] Imperceptible thickness, typically in the range of 1 nm of a monolayer 2D material, makes it intrinsically flexible, yet soft and adaptable, allowing it to conform to arbitrary shapes.[5] Typically, the single atomic layers of 2D materials are highly transparent in the visible wavelength region (97% transparency in the case of graphene).[6] These properties make 2D-TMDs generally attractive as fundamental components of future wearable bioelectronics.

Platinum is one of the most stable metals from the group 10 materials and is considered a noble metal (along with gold). It is well known for its resistance to corrosion,[7] catalytic properties,[8] and biocompatibility.[9] Pt is frequently used in fundamental research and industrial applications as an electrode, catalyst, etc. Consequently, Pt-based TMDs (Pt-TMDs) were theoretically expected and have since been proven experimentally to have good air- and water- stability,[10] compared to other TMDs. Pt-TMDs typically feature thickness-dependent semiconductor-to-metal transition,[4,11] ambient stability,[10] and low synthesis temperature.[12–14] Such properties position them as top candidates for the development of wearable bioelectronics devices.

Out of the different Pt-TMDs, two have recently attracted immense interest owing to their unique thickness-mediated change in electronic properties: platinum diselenide ($PtSe_2$) and platinum ditelluride ($PtTe_2$). Monolayer $PtSe_2$ is a semiconductor with an indirect bandgap of 1.25eV, which transforms into a metallic character with an increased number of layers.[4,15] $PtTe_2$, however, has a stronger interaction between the interlayer chalcogen atoms (Te), making the out-of-plane valence bond broader and overlapping, leading to its increased metallic properties even with small layer numbers. Consequently, $PtTe_2$ has metallic properties.[15–17]

An advantage of $PtSe_2$ and $PtTe_2$ compared to other TMDs is their low-temperature growth requirements. The thermal assisted conversion method allows for the growth of high-quality $PtSe_2$[4,12,14] and $PtTe_2$ [18,19] films on a large scale and at temperatures as low as 400°C, making it compatible with polymeric substrates.[20] The direct growth of 2D TMDs on ultrathin polyimide allows the fabrication of flexible materials with unique electronic properties that have high prospects of being the future building blocks of next-generation wearable devices. $PtSe_2$ and $PtTe_2$ are considered promising for numerous bioelectronic applications, yet none has been practically realized so far.

In this present work, we show how $PtSe_2$ and $PtTe_2$ can be used to develop multipurpose and reusable electronic tattoos with diverse applications. In terms of electrical conductivity, skin-contact, and electrochemical impedance, $PtTe_2$ outperforms graphene electronic tattoos,[21] and performs on par or even better than classical Ag/AgCl gel electrodes (despite the ~12 times smaller area). Skin impedance based electrode characterization data is supplemented with the electrical impedance spectroscopy (EIS) measurements, both suggesting that ultrathin $PtTe_2$ is remarkable material to be used as an electrode for future bio- and neuro-interfaces. In the present work, we showcase several applications of Pt-TMDs for numerous wearable healthcare applications. One of them is measuring electromyograms (EMG, muscle contractions) for building next-generation human-machine interfaces. Recording electrooculograms (EOG), eye movement-related electrical activity, is demonstrated as the prototype application for the imperceptible skin-worn electronic tattoos. By placing electrodes on top, bottom, and to the sides of a subject's eye, we built a soft and ultrathin interface that can capture eye movements, hence can be further used for robotic interfaces. Continuous and fault-free monitoring of electrocardiogram (ECG, heartbeat) and electroencephalogram (EEG, brainwaves) are essential aspects of hospital-based biopotential monitoring systems. Here, we performed exactly the same tasks with $PtTe_2$ and $PtSe_2$ tattoos, paving the way towards personalized and mobile healthcare. Body temperature monitoring is another example of an essential vital biomarker. We have found experimentally that the body temperature can be precisely monitored utilizing $PtSe_2$ and $PtTe_2$ sensors that remarkably show opposite signs in the measured temperature electrical coefficient (TEC).

## Results and Discussion

$PtSe_2$ and $PtTe_2$ used in this work are grown by thermal assisted conversion (TAC) based chemical vapor deposition (CVD) process. The growth is performed at moderately low temperatures, at about 400 °C, allowing us to grow the material directly over temperature-sensitive polymers.[13] A classic polymer commonly used in bioelectronic applications is Kapton, known for its mechanical and temperature stability. In this work, we use two kinds of Pt-TMD tattoos: some are grown directly over Kapton, others are grown on $SiO_2$/Si wafers, and then transferred onto an ultrathin temporary-tattoo-form utilizing mechanical support of 200 nm thick polymethyl methacrylate (PMMA) layer.[21] The TAC process requires platinum to be pre-deposited on the samples' surface (see Figure 1a and Figure S1a). We use 6 nm thick Pt as the material source for Kapton structures, and 3 nm thick Pt as the source for $SiO_2$/Si growth material. The 6 nm thick Pt samples, post-CVD growth result in ~25±3 nm thick multilayer Pt-TMD that features multiple out-of-plane growth planes (see Supporting Information for details). The 3 nm thick samples typically result in ~15±3 nm thick layered dichalcogenide, with a more uniform in-plane layered arrangement, which has been deemed beneficial for PMMA-supported tattoo material. After the CVD growth is performed (Figure 1b), the Kapton-based samples are ready for final processing. In contrast, the Si-based ones require further additional transfer steps (See Methods for details). The PMMA- and Kapton-based Pt-TMD tattoos are finally cut into desired shapes by means of a cutter-plotter tool (Silhouette Cameo). The PMMA based tattoos are just slightly scribed, while the Kapton based samples are cut through their thickness. The Kapton samples are also fixed onto temperature release tape (TRT) to allow precise and fault-free cutting (see Figure 1d). The final devices are of arbitrary shape and either supported by a 25 μm thick Kapton or a 200 nm thick PMMA (Figure 1e and Figure S1i).

To confirm the quality of the as-grown $PtSe_2$ and $PtTe_2$ layers, we carried out structural characterization employing X-ray photoelectron spectroscopy (XPS) and Raman spectroscopy. Figure 2a shows the Pt4f and Se3d core-level spectra obtained from the Pt-deposited substrates after undergoing thermal assisted conversion to $PtSe_2$ layers in the CVD furnace. The Pt4f spectra shows no elemental Pt signal, indicating that the entire spectrum can be assigned to $Pt^{4+}$ that corresponds to $PtSe_2$. Similarly, the Se3d spectra represents $PtSe_2$ whose spin-up state (j = 2 +1/2) is located at ~ 54 eV.[13] The XPS profile in Figure 2b (left) shows the core-level peaks of Pt4f at ~ 72.2 eV and ~ 75.1 eV, assigned to the doublet $4f_{7/2}$ and $4f_{5/2}$ of $PtTe_2$. The XPS profile reported in Figure 2b (right) shows the Te-3d core-level spectra with peaks at ~ 576 eV and ~ 586 eV that correspond to the Te (iv) species as well as the additional peaks at ~ 572.6 eV and ~ 582.9 eV, corresponding to Te (0). The peaks obtained from the Pt-4f and Te-3d core-level spectra of $PtTe_2$ are also consistent with previous reports.[22] Additionally, Raman spectra were collected from the PtTe$_2$ and PtSe$_2$ layers directly grown on SiO$_2$/Si substrates. The Raman profiles in Figure 2c shows two distinct peaks corresponding to vibration modes for PtSe$_2$ and PtTe$_2$ layers. The PtSe$_2$ samples exhibit two dominant peaks, E$_g$ mode at 175 cm$^{-1}$ and A$_{1g}$ mode at 205 cm$^{-1}$ correspond to the in-plane and out-of-plane vibration motions of the Se atoms, respectively.[23] Similarly, the Raman spectra collected from PtTe$_2$ shows peaks at ~110 cm$^{-1}$ and 157 cm$^{-1}$ correspond to E$_g$ and A$_{1g}$ modes.[24,25]

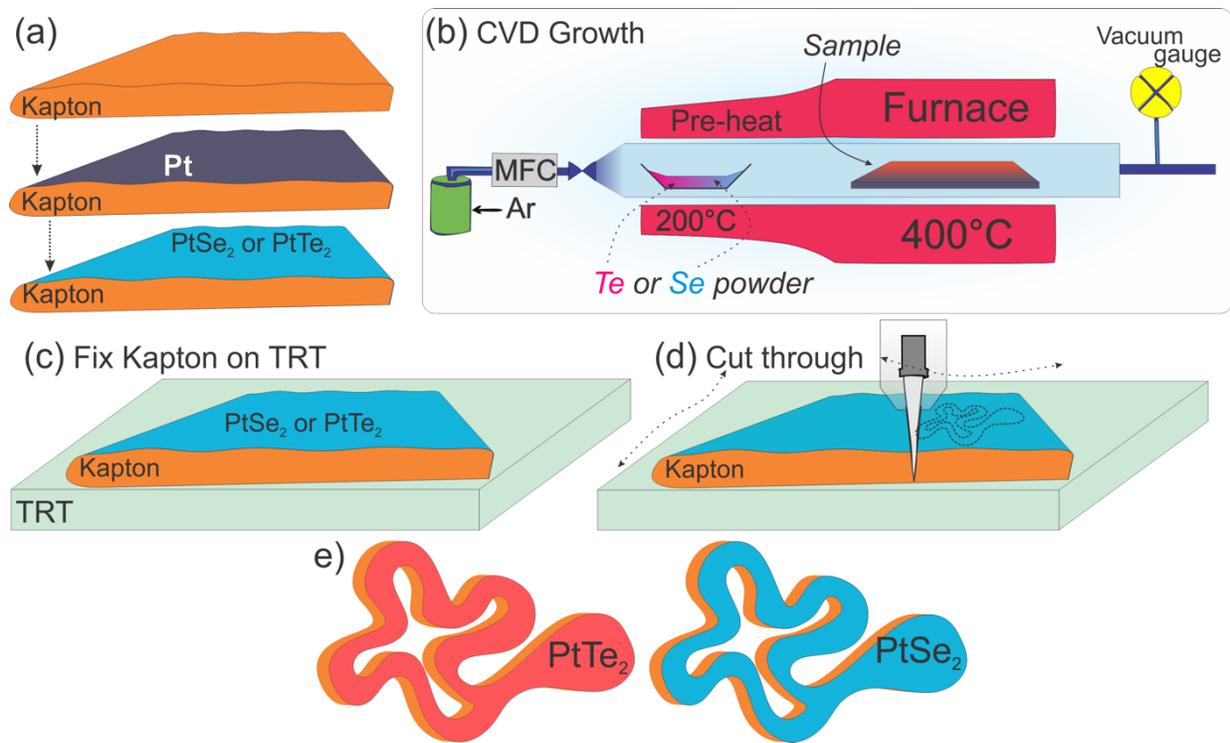

Figure 1. Schematic overview of Kapton based Pt-TMD tattoo design & fabrication flow. (a) Evaporation of thin Pt on top of the Kapton film, followed by TAC conversion into Pt-TMD. (b) Schematic of the TAC CVD process. (c) Post CVD growth, the Pt-TMD/Kapton sample is fixed on top of a TRT. (d) The mechanical patterning process of the Pt-TMDs grown on Kapton film. (e) Schematics of the final PtSe$_2$ and PtTe$_2$ tattoos supported by Kapton.

As one can see from the photographs (e.g., Figure 3a and Figure S17), the tattoos are not entirely transparent. This is mainly due to the number of TMD layers, and the optical transparency can be greatly improved if TMDs of smaller thickness are used. Having confirmed the growth of $PtSe_2$ and $PtTe_2$ layers, we investigated the thickness-dependent optical properties using UV-Vis spectroscopy. Figure 2d shows the optical transmittance (at 550 nm) along with sheet resistance and of the $PtSe_2$ and $PtTe_2$ layers with respect to the material's thickness (See Figures S3 and S4 for the full dataset). $PtSe_2$ films become more semiconducting as the thickness reduces, according to recent reports on $PtSe_2$-based electrical devices.[14,26] The corresponding sheet resistance of the $PtSe_2$ layers reduces from ~100 kΩ/□ to ~1 kΩ/□ as the thickness increases. In contrast to $PtSe_2$ films, $PtTe_2$ retain their metallic transport properties at varying thickness, which is in accordance with recent reports.[25,27] The sheet resistance of $PtTe_2$ ranges from ~350 Ω/□ at 1 nm, down to as low as ~31 Ω/□ at 6 nm. $PtSe_2$ and $PtTe_2$ layers with Pt seed layer thickness of ~1 nm have optical transmittance of ~50%, with a noticeable increase in the optical absorbance with increased thickness, similar to the observation in other 2D TMDs. The multilayered $PtSe_2$ and $PtTe_2$ materials used in this study remain ultrathin with a post-growth thickness of less than 30 nm (see Figure S4 for details).

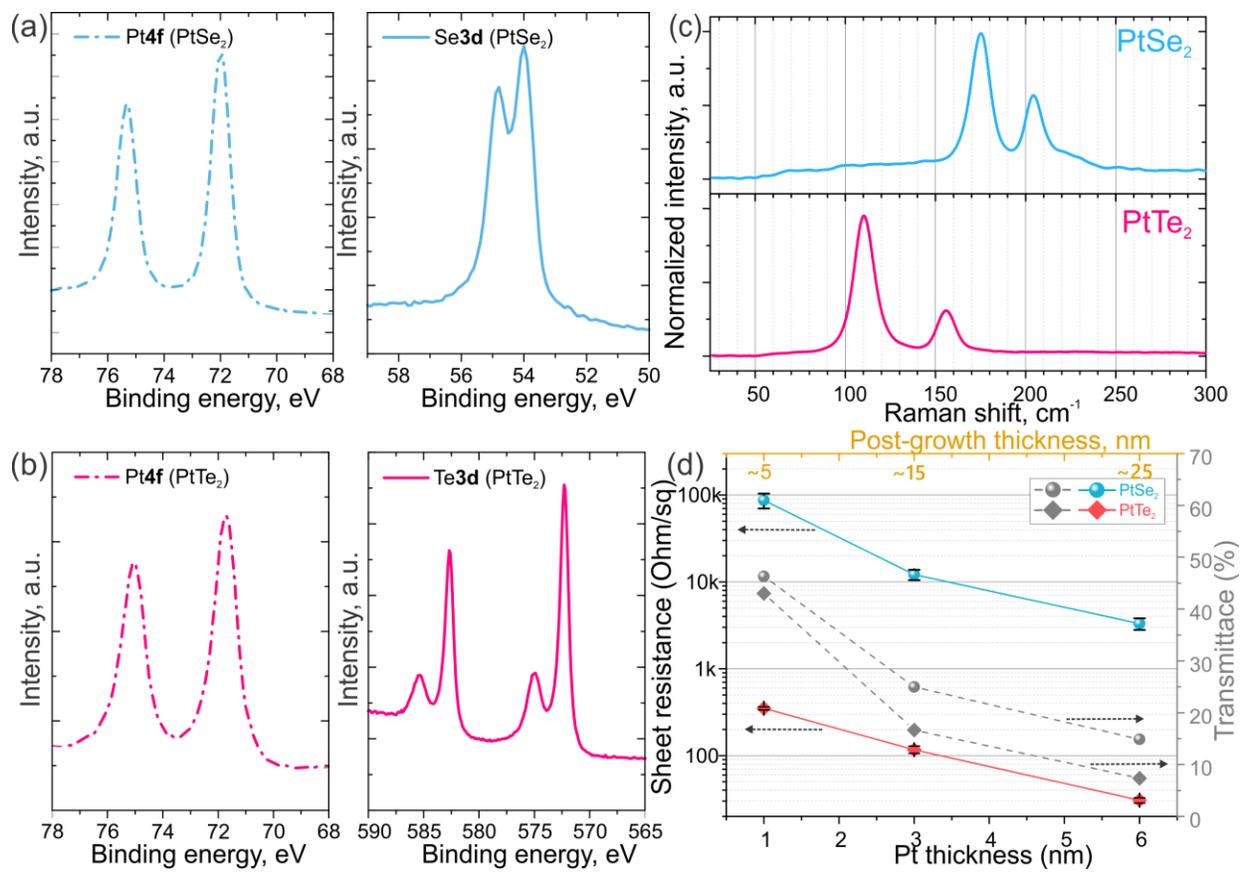

Figure 2. (a) XPS characterization of Pt4f and Se3d core levels obtained from PtSe$_2$ layers on SiO$_2$/Si substrates. (b) XPS characterization of Pt4f and Te3d core levels obtained from PtTe$_2$ layers on SiO$_2$/Si substrates. (c) Raman spectra of as-grown PtSe$_2$ and PtTe$_2$ layers. (d) Sheet resistance and optical transmittance (at 550 nm) of PtSe$_2$ and PtTe$_2$ as a function of its thickness, pre-growth, and post-growth.

Once fabricated, both kinds of tattoos can be easily placed over the skin for characterization and electrophysiology. The temporary tattoo electrodes, supported by the PMMA, are transferred similarly to the graphene electronic tattoos.[21] PMMA provides a more imperceptible and intimate contact with the skin. As can be seen, the tattoos come in a very secure contact with the skin, surviving motion, bending, and stretching. On the other hand, the Kapton-based structures can be simply held in place via a piece of Tegaderm or kind removal silicone tape (KRST), making them reusable. Pictures of both types of Pt-TMD tattoos supported by PMMA and Kapton on skin are shown in Figure 3a, and Figure S17.

To examine the electrode-to-skin impedance of Pt-TMDs and compare them to the state-of-the-art, we fabricated tattoos made of bare platinum, bare gold, $PtSe_2$, and $PtTe_2$ over polyimide substrates. The tattoos have been manufactured as explained above, with at least 5 pairs of each kind used to study the electrode-skin impedance. Graphene tattoos have been fabricated as explained elsewhere.[21] Considering that the actual tissue impedance is in the range of several hundreds of $m\Omega$ to dozens of $\Omega$,[28] the skin impedance value measured between the electrode pair placed over the skin will give us the electrode-skin impedance. This property is commonly considered as a figure-of-merit of the wearable electrodes.[29–31] A typical Pt-TMD-skin impedance plot can be seen in Figure 3b, featuring the regular 1/f frequency dependency.

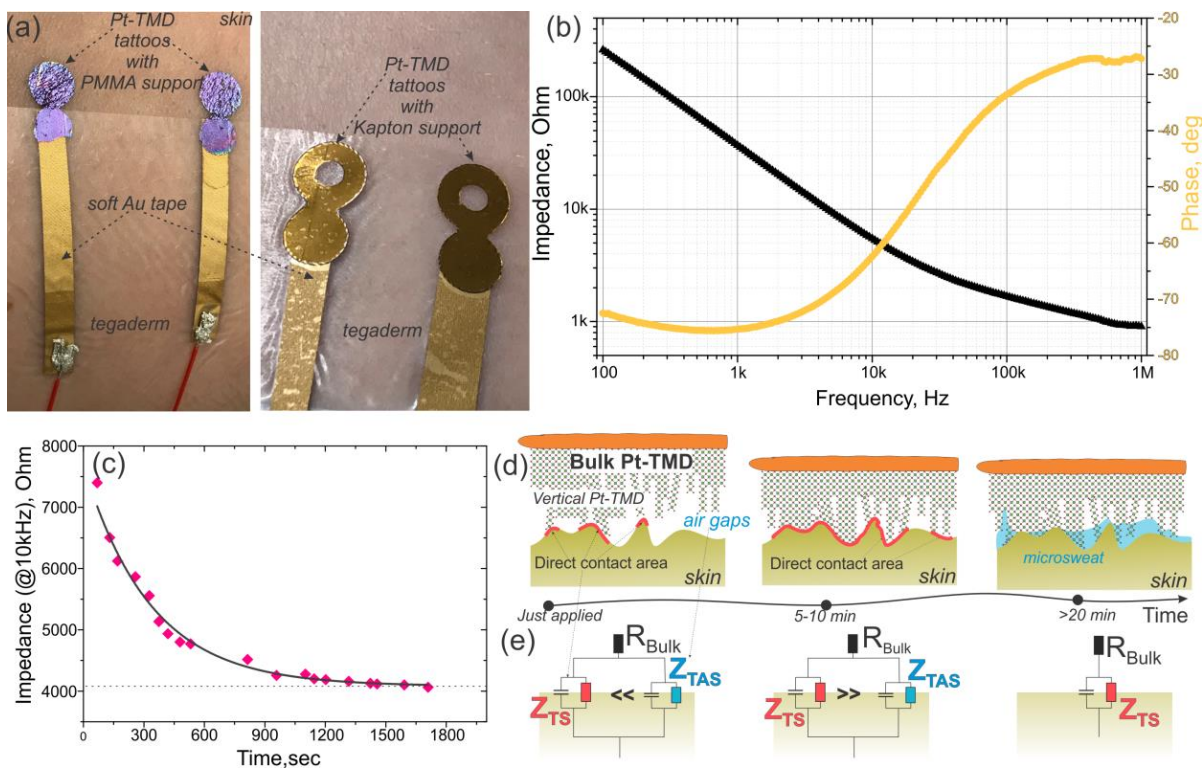

Figure 3. (a) Pictures of Pt-TMD tattoos, with PMMA (left) and Kapton (right) polymeric support and built up electrical contacts. (b) Typical Bode diagram of the $PtTe_2$-Skin impedance. (c) The temporal exponential decay of the contact impedance (@10 kHz) upon applying PtTe$_2$ tattoos on the skin. (d-e) The proposed schematic of the time-related dynamic changes to Pt-TMD-skin interface with suggested equivalent circuits. $Z_{TS}$ and $Z_{TAS}$ represent direct tattoo-skin and indirect tattoo-air-skin complex impedances, respectively.

A peculiar dynamics was observed for the case of Pt-TMD based tattoos, that just upon placement onto the skin, the electrodes feature relatively high impedance (~10 kΩ @10kHz) that rapidly reduces over a time period of about 15 mins to reach a steady-state value of ~4 kΩ (Figure 3c). We believe this might be related to the structural properties of the thick Pt-TMDs. As reported previously, CVD growth of thicker (6 nm seed layer) Pt-TMDs results in multiple 2D material domains merging together (see Figure 3d).[14] This would result in firstly uneven height distribution and nanoscale rough surface on the tattoo. When placed onto the skin, the rough nano-surface of Pt-TMDs in combination with the microporous structure of skin results in a high-impedance condition due to the presence of trapped air, reducing the overall direct contact area. Over time, the tattoos forge a more intimate contact with skin, reducing the air gaps, and establishing a closer connection. Moreover, we hypothesize that upon certain conditions (e.g., when Tegaderm fixes the Kapton-based tattoos), the macroscopic sweat and moisture gets trapped between the skin and Pt-TMD, further improving the contact impedance.[29] To corroborate this hypothesis, we performed curve-fitting of the frequency-dependent impedance data measured over time upon application over the skin (20 points, with ~90 sec intervals). As given in Figure 3e, the proposed equivalent circuits are impossible to fit since the number of variables is too large, and the model can converge at any arbitrary value. However, upon fitting the impedance with a simple Randles circuit[32] (see Figure S5), the resulting fit is profoundly unstable in the first 900 seconds after tattoo application, yet yielding relevant and reliable values after 900 seconds. The results indicate that the simplified Randles circuit with low ohmic access resistance (<400 $\Omega$), low contact resistance (~200 k$\Omega$), imperfect (a=0.85) capacitor, and a constant phase element (CPE, ~8.5 nF) is a reasonable estimation of the interface after ~900 sec upon tattoo application, but is a poor estimation of the interface during the initial transitory contact-forming stage.

To ensure that the figure-of-merit values of Pt-TMD-to-skin impedance are most representative, we let the tattoos stay on the skin for at least 20 minutes before performing the measurements. Multiple batches of Pt-TMDs have been utilized in order to make sure that the data reported represents the median material behavior and not a random sample. At least four pairs (N>8) of each tattoo kind were used to gather the statistically relevant data. The impedance (@10 kHz) distribution of the PtSe$_2$, PtTe$_2$, and Pt samples is shown in Figure 4a. It is clear from the figure and the numbers that the PtTe$_2$ offers the lowest impedance, averaging 4.94±1.61 k$\Omega$. PtSe$_2$, in comparison, averages at 7.24±2.56 k$\Omega$, which is slightly worse than bare Pt with 6.75±0.59 k$\Omega$. The full statistical distribution of the electrode-skin impedance values for PtSe$_2$ and PtTe$_2$ is given in Figure S6. Electrode dimension is an important parameter to keep in mind since the classical electrical circuit model states that the electrode-skin impedance is inversely proportional to the electrode's area.[33] Therefore, all electrodes utilized in this work (except the Ag/AgCl gel ones) are made with a total area of 25 mm$^2$. The Ag/AgCl electrodes cannot be cut smaller, and hence we have trimmed them down to only ~100 mm$^2$ for electrode-skin measurements. Still, they are kept at their original size ~300 mm$^2$ for most of the electrophysiological measurements. Regardless of such four-time difference in surface area, some of the highest quality PtTe$_2$ tattoos have shown better performance than the medical-grade Ag/AgCl gel electrodes. The sheet resistance of polymer-supported Pt-TMDs was measured by placing a strip of the material out of the same growth batch onto the Ecoflex-covered glass with four soft conductive gold electrodes. Ecoflex is used here to reproduce skin-like texture and softness. The 4-point transfer length measurement (TLM) was performed to estimate sheet resistance. The values reported are averages from at least 2-3 strips of each kind. The sheet resistance of $PtTe_2$ tattoos is exceptionally low, averaging at $56\pm49$ $\Omega/\square$, with the best-reported value of only $13$ $\Omega/\square$. In comparison, $PtSe_2$ tattoos hit about $1.11\pm2.37$ $k\Omega/\square$, which is almost 20 times higher. This can be attributed to the different nature of these two 2D materials: $PtTe_2$ is highly metallic, while $PtSe_2$, even in its bulk form, is semimetallic.

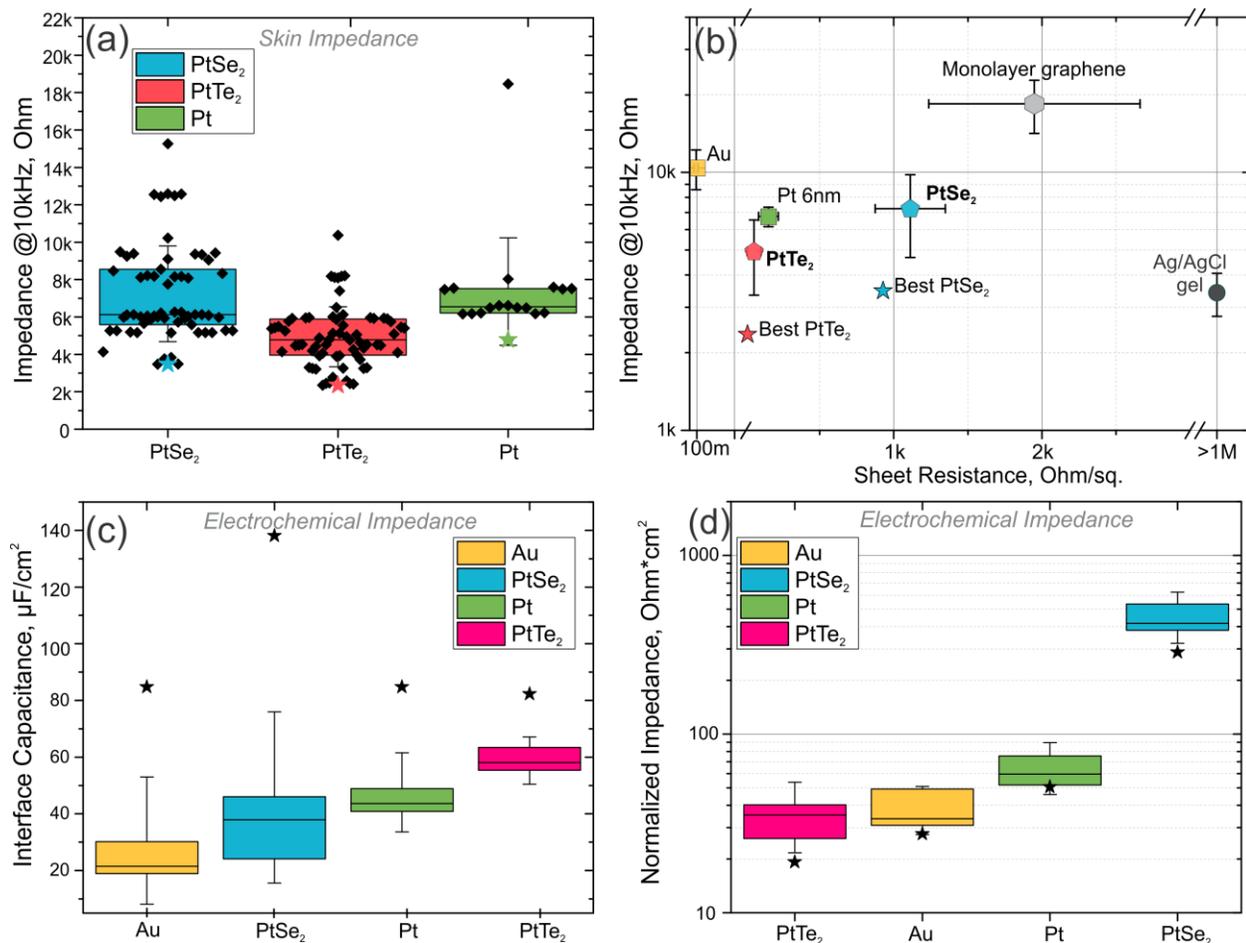

Figure 4. (a) Statistical distribution of the electrode-skin impedance values for $PtSe_2$ (6 nm, blue), $PtTe_2$ (6nm, pink), and bare Pt (6 nm, green). (b) Comparison scatter graph of both skin impedance and sheet resistance of Au (100nm, yellow), Pt (6nm, green), $PtTe_2$ and $PtSe_2$, monolayer graphene tattoos (light grey), and Ag/AgCl gel (dark grey). Impedance is reported at 10 kHz frequency and is averaged from N>8 electrodes. The sheet resistance is averaged from N>3 TLM measurements and different devices. (c-d) Electrochemical impedance spectroscopy data of the Pt-TMD's: interface capacitance and normalized impedance (@ 1kHz) correspondingly. In (a), (c), and (d), the box represents 25% and 75% with a mean, outliers are ±SD, and stars represent the best examples.

The overall plot of merit that combines sheet resistance and skin impedance data for PtSe$_2$, PtTe$_2$, Pt, gold, Ag/AgCl, and even monolayer graphene for comparison is shown in Figure 4b. The major trend noticeable from the plot is that PtTe$_2$ is much more promising for wearable applications than PtSe$_2$. Both PtSe$_2$ and PtTe$_2$ yield better skin impedance than bare Pt. PtTe$_2$outperforms almost every other competitor material, including bare gold, and is on-par with the medical-grade Ag/AgCl electrodes. Significantly, both PtSe$_2$ and PtTe$_2$ outperform monolayer graphene tattoos in terms of both electrical properties, which is a significant improvement, considering that graphene tattoos typically feature an advanced adhesion to skin[21,34].

To supplement and corroborate the skin impedance values, we have performed electrochemical impedance spectroscopy (EIS) study of these materials. For EIS experiments, the Kapton-supported TMD tattoos and control samples were made with varied areas to achieve scalable results. The EIS measurements were taken in a three-electrode configuration; see details in the Methods section. Once measured, the EIS data was evaluated and fitted with an equivalent circuit, extracting the interface capacitance, and the electrode impedance (@1 kHz). The area-normalized interface capacitance and area-normalized electrode impedance of PtSe$_2$, PtTe$_2$, Au, and Pt are shown in Figures 4c-d. Similar to the skin impedance, PtTe$_2$ exhibits the most outstanding performance, featuring exceptionally high interface capacitance of 62.2±8.2 µF/cm$^2$ and smallest normalized impedance at 1 kHz of 35.4±18.6 Ω•cm$^2$. PtSe$_2$, as expected, is showing a lower interface capacitance of 45.8±30.2 $\mu F/cm^2$ and higher normalized impedance at 1kHz of 472±151 $\Omega \cdot cm^2$. To put these values in perspective, the normalized impedance at 1 kHz of monolayer graphene is averaging at 74±173 $\Omega \cdot cm^2$, [35] which is much higher than PtTe$_2$. The high interface capacitance and small normalized impedance are typically good signs of material's performance and its usability for building advanced microelectrode arrays (MEAs) for *in vitro* and *in vivo* applications. It is essential to mention here that the electrodes used for EIS in this work are macro-sized compared to micro-sized electrodes typically used in MEAs. We expect the overall interface capacitance of the micro-sized Pt-TMDs to improve when scaled down. Furthermore, considering the 2D nature of PtTe$_2$, its flexibility and stability, it is certainly a suitable bioelectronic material, and we expect the works of PtTe$_2$ based MEAs or *in vivo* probes to follow soon.

To begin with the electrophysiological measurements, TMD tattoos were laminated on a subject's chest to measure ECG. At this moment and further, all electrophysiological measurements were performed by means of a simple off-the-shelf open-source board – *Ganglion* from OpenBCI. In the case of ECG recordings, one pair of Pt-TMD tattoos is placed on the chest as schematically shown in Figure 5a. Another couple of Ag/AgCl gel electrodes is placed nearby with similar spacing to provide a direct signal comparison (see Figure 5a). A reference ground-driven electrode is placed over the lower right of the abdomen. The signals are sampled at 200 Hz, and later filtered with a 60 Hz notch filter to remove the power line noise. The ECG signals measured with Ag/AgCl and PtTe$_2$ are shown in Figure 5b-c with clearly present characteristic ECG peaks (P, Q, R, S, T).[36] The signal to noise ratio (SNR) of PtTe$_2$ tattoo based recordings was as high as 84±6, while the nearby Ag/AgCl gel electrodes yielded 61±5. Similar ECG signal recordings with PtSe$_2$ tattoo are also shown in Figure S7. PtSe$_2$ tattoos' performance is slightly lower, yet the SNR of 44±4 is high enough to provide relevant signal information and shape, especially considering the 12-fold difference in the electrode size. Besides, ECG can be recorded in a less sophisticated way by placing two tattoos on opposite hands, e.g., forearms, and a single reference electrode near a bone, e.g., elbow. Such data is shown in Figure S8 featuring a clear and high amplitude ECG signal having the essential large P, Q, R, S, and T domains.

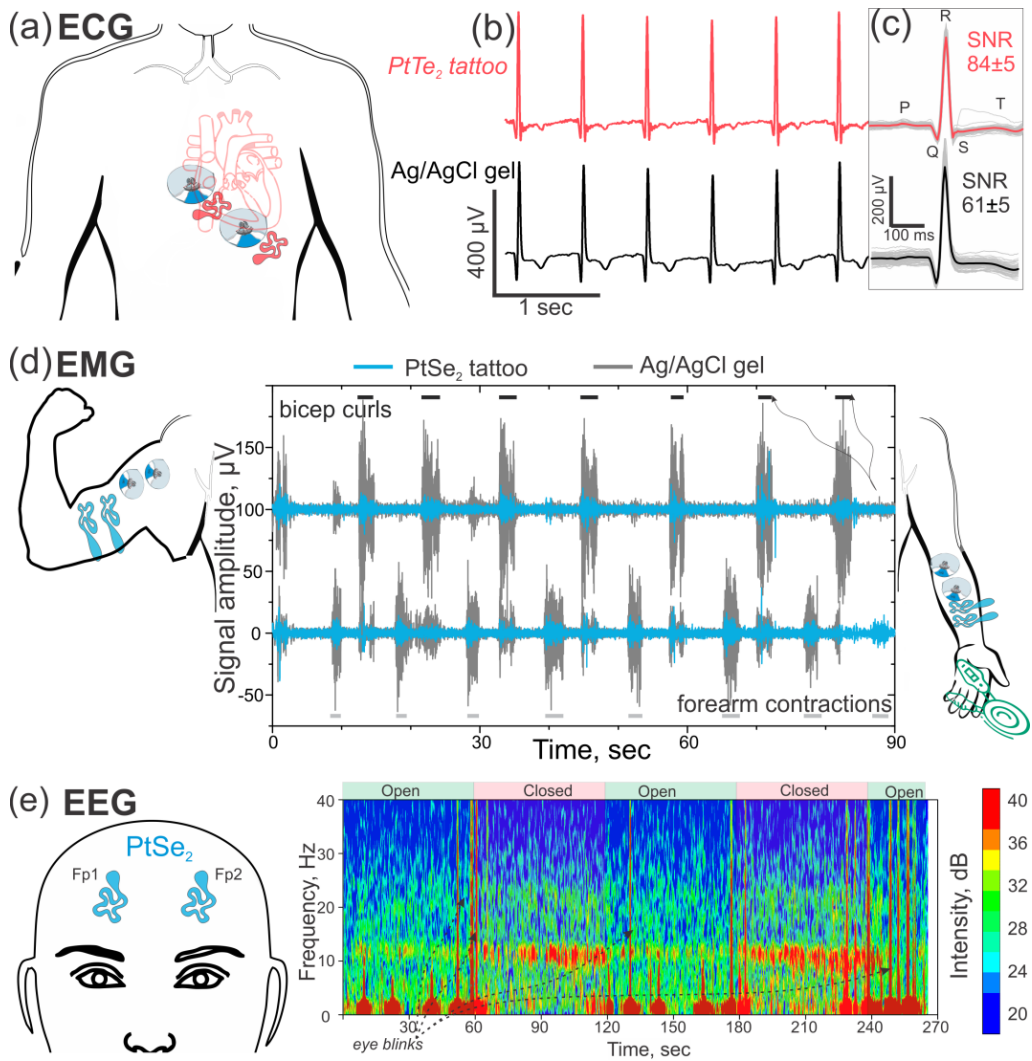

Figure 5. Wearable electrophysiological recordings with Pt-TMDs. (a) Schematic of the electrode placement for the ECG measurement. (b) The 5-second-long ECG timetrace from $PtTe_2$ tattoos and Ag/AgCl gel electrodes. (c) The average ECG signal shapes, amplitudes, and average SNRs for $PtTe_2$ tattoo and Ag/AgCl gel electrode. (d) Schematic of electrode placement for EMG, and EMG recordings from a human's bicep (top), and forearm (bottom) muscle contractions recorded via the Ag/Ag gel electrodes and PtSe$_2$ tattoos simultaneously. The horizontal bars on the top and bottom mark the events and durations of the bicep curls and forearm contractions correspondingly. (e) The electrode placement for EEG measurements (left) and resulting spectrogram (right) from one of the PtSe$_2$ tattoos placed on the Fp1 and Fp2 locations on the subject's forehead.

EMG is a distinct, powerful, and useful electrical signal that corresponds to the muscles' physical contractions within the body.[37] When a pair of electrodes biased in a differential manner is placed over a muscle, the muscular contractions will generate a net electrical potential. Typically, the larger the distance between the two electrodes, the larger is the differential signal recorded from the pair. In order to record useful EMG signals, we placed a pair of Pt-TMD tattoos on a forearm and another pair of tattoos on a bicep, along with the pairs of Ag/AgCl electrodes for direct comparison (see schematics on Figure 5d). During a series of forearm contractions (mechanical expander) and bicep curls, we clearly distinguished those movements via simple monitoring of the EMG activity and the signal's power. One can see that the signal power from gel electrodes is higher compared to Pt-TMDs, which is due to 12-times larger electrode area as well as the spacing between the electrodes is almost 5-times higher. Nonetheless, as it can be seen from Figure 5d - for PtSe$_2$, and Figure S9 - for PtTe$_2$, we are able to optimally recognize both: movements of the forearm and bicep, as well as differentiate between them. On the brighter side, Pt-TMD electrodes feature lower noise, essentially displaying higher signal specificity, equalizing the performance. As it can be seen from the timetrace of consecutive bicep and forearm movements (see Figure 5d), often, the muscular movement associated with a bicep curl also results in slight forearm muscle contractions, which is always picked up by the Ag/AgCl gel electrodes. These non-specific signals usually obstruct from making clear and decisive interfacing, and as it is clear from the results provided, Pt-TMD tattoos suffer much less from this problem.

The low contact electrode-skin impedance displayed by Pt-TMDs is essential for most electrophysiological measurements, but especially for EEG. EEG signals are electrical signals associated with the ever-active electrical activities inside the brain.[38,39] There is a broad scientific community working on EEG and its applications, mostly studying the correlations between EEG signals and the underlying brain functions. Developing a tattoo system that can be used for EEG recordings (a) reliably, (b) with intimate contact and signal quality, (c) reusably, is a high-impact task. To measure EEG signals, a couple of Pt-TMD tattoos were placed on the subject's forehead, precisely onto the Fp1 and Fp2 locations,[40] and sampled separately with a common ground electrode connected to an earlobe. The EEG recordings were typically collected over a period of 120-240 seconds and featured specific patterns of subjects' activities - their eyes being open or closed. Post applying the short-time-Fourier-transformation (STFT) onto the recorded signals, the output spectrogram, as shown in Figure 5e, reveals the heatmap of signal frequency components and their power levels (in dB), distributed over the entire timetrace. It is well known that EEG, recorded from the forehead, should show prevailing alpha waves (8-13Hz) when the subject's eyes are closed, especially at the Fp1 and Fp2 locations.[41] As seen from Figure 5e and Figures S12, the spectrogram patterns pick up the alpha waves when 'eyes closed' with $PtSe_2$ and $PtTe_2$ tattoos, and Ag/AgCl gel electrodes. To specifically show the statistical presence of alpha rhythms, we have averaged and plotted the spectral power density from gel, $PtSe_2$, and $PtTe_2$ electrodes with open (30 sec) and closed (30 sec) eye intervals. As seen from the FFT data (see Figure S13), there is a significant power density peak at around 8-13 Hz that corresponds to the alpha waves when the 'eyes closed', yet the peak is absent in the case of 'eyes open'.

To showcase a human-machine interface enabled by the Pt-TMD tattoos, we measured electrical signals associated with the eye movements,[42] which could easily be later channeled into a computer, robot, or machine.[43,44] EOGs are very similar to the EMG signals but are associated with the differences in polarization of the eye's cornea and retina.[45] To record EOG signals, we have placed four Pt-TMDs on a subject's face: one pair goes above and below the left eye, one tattoo on the left side of the left eye, and the last tattoo on the right side of the right eye (see Figure 6a for details). The tattoos placed on the sides of the eyes are connected differentially into a single channel, and the electrodes above and below the left eye are connected differentially to the other channel. When the subject is looking into a particular direction, specific cornea-retina de-polarization events create electric potential differences associated with the movement. As seen from the timetrace recordings (see Figure 6b), gazing left and right results in the distinct change of the appropriate channel's EOG signal. Similarly, gazing up and down results in a specific pattern recorded by another pair of electrodes, see Figure 6c. The EOG signals, as shown in Figure S14, for example, are very sharp and associated with rapid and repetitive eye movements without a distinct baseline. When slower gazing experiments are performed, a clear differentiation can be made between various gazing patterns (e.g., up-down, up-center, left-center, center-right, etc.). There is, however, a certain 'overshoot' signal that can be seen in some recordings (see Figure 6b-c) when the eyes are returning from a peripheral into straight positions. However, these 'overshoots' are directional and help distinguish the change of sight direction in dynamics. Figure 6d shows the results of an elaborate, non-routine experiment when the subject was instructed to look into the direction as conducted by the experimenter, online and without any warning. It is visible that the patterns for looking down, up, left, and right can be discretely identified. Such a pattern is readily distinguishable, allowing us to differentiate between arbitrary directions. Eye blinks do cause some troubles and create certain events, mainly in the *up-down* channel, yet they can be later post-processed and removed.[46]

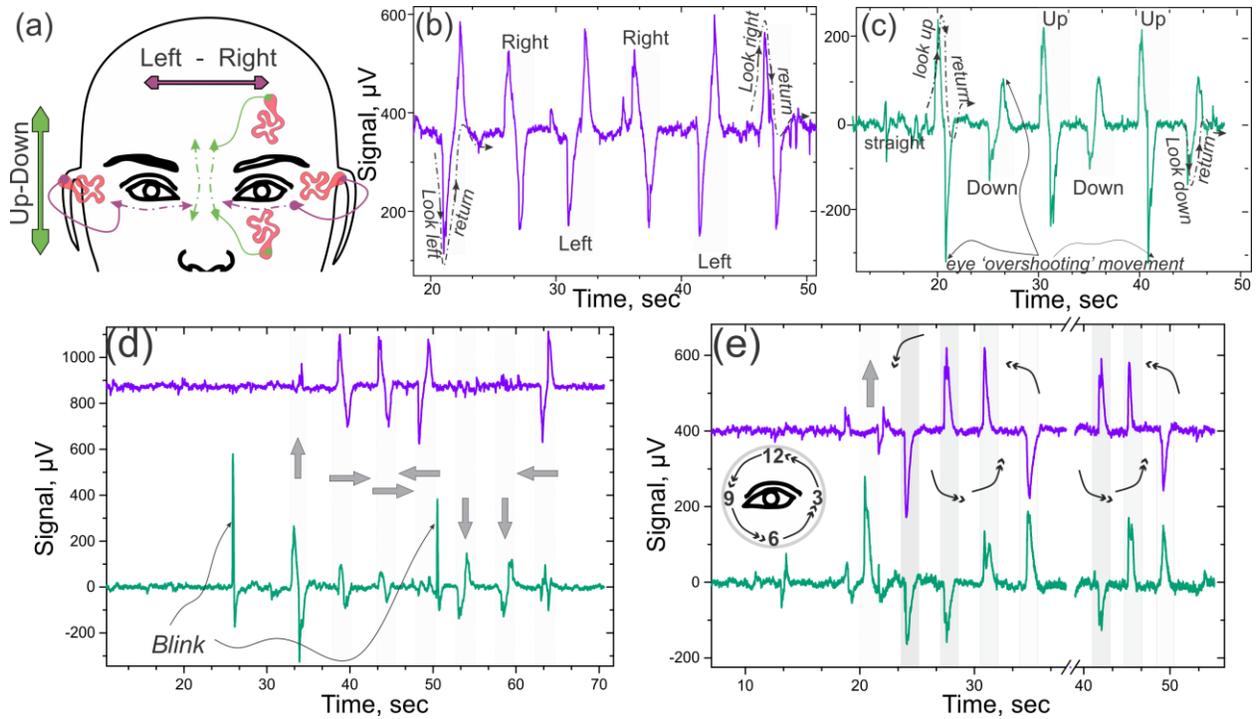

Figure 6. (a) Schematic of the electrode placement for the EOG experiments performed with PtTe$_2$ tattoos. (b) The EOG signals corresponding to the differential channel recording left-right eye movements (violet here and after). (c) The EOG signals corresponding to the differential channel recording up-down eye movements (green here and after). (d) The EOG recordings from both channels when the subject is gazing up, right, right, left, down, down, and left consecutively with 5 seconds dwell time. (e) The EOG recordings from both channels when the subject is looking counterclockwise with 4 distinct stops: at '12', '9'. '6', and '3' o'clock. Straight and angular arrows in (d) and (e) correspondingly, show the direction of sight.

In the final EOG experiment, the subject was instructed to perform more sophisticated eye movements that are not just fixed to one degree of freedom (straight horizontal or straight vertical) as before. Here, the subject performed circular eye movements (like looking over a watch) counterclockwise with four distinct stops: at 12, 9, 6, and 3 o'clock positions. Astonishingly, those movements resemble a combination of the afore-mentioned two-degree movements. For example, when moving from 12 to 9, the recorded signal pattern is a combination of the signals corresponding to "left" and "down" (see Figure 6e for details). The results clearly indicate that the TMD tattoos can constitute a technology that does advanced eye-tracking for human-machine interfaces.

In addition to recording human electrophysiological signals, the Pt-TMD tattoos can also measure skin temperature. Continuous and precise human body temperature monitoring is an important feature for next-generation wearables. Tracking and monitoring the body core temperature changes is essential to provide an early response to possible viral infection or disease. With Pt-TMD tattoos, we have found the task of temperature monitoring very intriguing. Here, the two dichalcogenides of choice turned out to have opposite signs of temperature electrical coefficients (TEC). The TECs were evaluated by measuring the tattoo's resistance while changing the environment's temperature and having a commercial thermocouple (TC) to control the actual surface temperature. As one can see from Figures S15-16, the TEC of $PtTe_2$ has a positive sign, and the alpha coefficient ranges between 0.0004 and 0.0020; while $PtSe_2$ features a negative TEC with the coefficient between -0.0010 and -0.0013. Here, $PtTe_2$ behaves like the majority of metals and increases its resistance with increased temperature. In $PtSe_2$, however, the resistance decreases with increased temperature featuring negative TEC, which is unexpected, since the $PtSe_2$ material used for this experiment is a highly conductive *semimetal*. Furthermore, the overall temperature response is linear over a broad temperature range tested, from 15°C and beyond 60°C, which is well within the human body temperature range. To validate Pt-TMD tattoos for measuring real-time skin temperature, we placed them onto a subject's forearm with a thermocouple nearby. After approximately 2 minutes of control measurement, an ice bag was brought in contact with the skin, held for a couple of minutes, then removed. The timetrace of PtSe$_2$'s resistance change and thermocouple's response can be seen in Figure S15b. As expected, upon cooling down, PtSe$_2$ resistance had increased. Moreover, it is evident that there is no lag in the response in comparison to commercial thermocouples, i.e., the slow decays of changes in resistance visible in Figure S15b are associated directly with the temperature profile and not to the materials' properties.

## Conclusions

In conclusion, we present in the work novel Pt-TMDs, namely PtSe$_2$ and PtTe$_2$ that can be used to build high quality advanced electronic tattoos and for wearable bioelectronic applications. Both PtSe$_2$ and PtTe$_2$, particularly the latter, possess high conductivity desirable for wearable and implantable electrodes. In terms of electrode-skin impedance, a figure-of-merit for wearable electrode systems, Pt-TMDs outperform graphene-based electronic tattoos, state-of-the-art metal tattoos, and medical-grade gel electrodes. Brainwaves, heart function, muscle activity, skin temperature, and even eye-tracking are shown to be accurately captured by means of Pt-TMD electronic tattoos. The eye-tracking is of particular interest since it applies to wearable human-machine interfaces. Remarkably, we have successfully used two of the pairs of PtTe$_2$ and PtSe$_2$ tattoos built on Kapton film for all electrophysiology modalities reported in this work, including EEG, EMG, ECG, and EOG, highlighting their reusability and versatility.

## Methods

**PtSe$_2$ and PtTe$_2$ growth**. The PtSe$_2$ and PtTe$_2$ layers were grown on the substrates of choice (SiO$_2$/Si, Kapton) using a two-step thermal assisted conversion process. Electron beam deposition (Temescal) was used to deposit Pt films of desired thickness at a rate of 0.1 Å/s on the substrates. The Pt-covered substrates are then placed in the middle zone of a CVD tube furnace with alumina boat containing tellurium (Te) or selenium (Se) placed at the upstream region for tellurization or selenization to obtain $PtTe_2$ or $PtSe_2$ layers, respectively. The CVD tube is pumped down to a low pressure of 25 mTorr followed by purging with Argon (Ar) gas at a high flow rate to remove the moisture and residual gases. For both $PtTe_2$ and $PtSe_2$, the furnace is heated to the growth temperature of 400°C and held there for 50 minutes under a constant flow of Ar. The furnace is left to cool down to room temperature naturally.

**Optical transmittance and water contact angle measurements**. The optical transmittance of $PtSe_2$ and $PtTe_2$ samples directly grown on optically transparent willow glass were characterized using ultraviolet-visible (UV-Vis) spectroscopy (Cary WinUV spectrophotometer) in the wavelength range 200–800 nm. Water contact angle measurements on $SiO_2$/Si substrates were carried out using a goniometer (Rame Hart 90-U3-PRO), and the same water volume was used to ensure uniformity of all measurements.

**Raman and XPS characterization**. X-ray photoelectron spectroscopy (XPS) characterization of $PtSe_2$ and $PtTe_2$ samples was carried out using a Thermo VG Scientific K-α system under ultrahigh vacuum conditions. The system has an energy resolution of ~0.5eV and 100W X-ray spot of 400 μm$^2$. The binding energy of C 1s, 284.5eV are used to calibrate the XPS peaks, and Voigt functions are used for the peak fitting. Raman spectroscopy was performed in a Renishaw inVia micro-Raman system. The excitation wavelength of 532 nm with an incident beam power of ~1 mW and exposure time of 10 s is used for Raman. A 3000 l/mm grating is used for $< 5\,cm^{-1}$ resolution.

**Electrophysiology data collection**. The EEG, EMG, ECG, and EOG signals were all recorded via the Ganglion Open-BCI board.[47] The board is a low-cost yet effective tool. The board itself is approximately 8x8cm in size and powered by a 3.3V-6V battery, and has a BLE module on board, making it freely portable. The ganglion board features four high-impedance differential inputs. The channels are sampled separately, using maximum 4 pairs of electrodes (one pair for each channel) for monitoring EMG, EOG, and ECG. In order to record EEG, the differential inputs can be connected to a single reference electrode, which is typically located on the earlobe. The board records data with a sampling rate of 200 Hz.

**Skin Impedance Measurements.** The electrode-skin impedance is measured via the Hioki LCR meter IM3536, which has a max frequency sweep range between 4 Hz up to 8 MHz. For our study, however, we narrowed the measurement window down to 10 Hz – 1 MHz. The measurements are performed in the constant voltage mode with 50 mV AC amplitude, without DC bias. Each data point is measured four times and averaged instrumentally. The frequency sweep is usually performed three times, each sweep taking approximately 90 seconds, and 10 seconds delay between sweeps. The measured values correspond to the electrode-skin impedance since the tissue's bio-impedance is much smaller when compared to the typical electrode-skin impedance.[48]

**Electrophysiology Data Analysis.** The data processing was performed with MATLAB, and the designed code is available on request. Raw ECG, EMG, ECG, and EOG data, as acquired by the OpenBCI board, were utilized for analysis. The recordings are stored in four separate channels, and the sample values are arranged in the form of vectors. The raw signals have a significant noise in the 60 Hz frequency region, and hence an appropriate 60 Hz notch filter was implemented before any further processing. To analyze the ECG signals, we performed a peak finding algorithm. From each peak found, we extracted its amplitude, which was further averaged to get the average signal (S) value. In order to find the noise, we extracted n regions of 200 ms duration between two spikes and found the median absolute deviation (MAD) of that timetrace. Twice MAD was considered the noise (N). The signal-to-noise ratio (SNR) is calculated accordingly by dividing signal amplitude by noise level. The EEG recordings were typically in the range of 120-240 seconds long. Two out of four OpenBCI channels correspond to the Fp1 and Fp2 locations' EEG measurements while the other two channels were turned off. First, a 60 Hz notch filter was implemented, followed by linear detrending to remove the baseline. Also, a lowpass filter with a cut-off frequency of 50 Hz was designed and implemented to further cut-off the high-frequency noise. As a crucial part of this analysis, short-time-Fourier-transform (STFT) was implemented with the Hamming window of 128 sample length and the FFT length of 512 samples. The STFT output as a matrix consisting of frequency and time samples, the corresponding Fourier coefficients, and the power spectral densities in dB units were stored separately and utilized for further plotting. The formulae utilized here can be found elsewhere.[49]

**Electrical Impedance Spectroscopy.** We utilized the Autolab PGSTAT 128N Potentiostat with NOVA software for data acquisition and preliminary processing for EIS characterization. A 1x PBS buffer solution was used as the standard electrolyte for the measurements, Ag/AgCl as a reference electrode, Pt wire as a counter electrode, and the material of interest (Pt, $PtSe_2$, $PtTe_2$, and Au) was connected as the working electrode. Appropriate frequency ranges for the plots were set, and measurements were taken. Areas of 20 $mm^2$, 24 $mm^2$, 30 $mm^2$, 35 $mm^2$, and 40 $mm^2$ of the 2D-TMD and reference materials were utilized in the EIS measurement runs.

**Temporary Tattoo Fabrication**. A layer of PMMA (950 A4) is spin-coated on top of Pt-TMD/$SiO_2$/Si at the rate of ~2500 rpm for 60 sec, and hard-baked at 200°C for 20 minutes, resulting in ~200 nm thick support layer. A TRT film (120°C release temperature, 3195MS Nitto Denko Revalpha) is carefully placed in contact with the PMMA, avoiding bubble formation. It is important to note here that prior to the Pt evaporation and TMD growth, the $SiO_2$/Si surface was treated with oxygen plasma for 5 minutes to improve the surface's hydrophilicity and demote the material's adhesion. The TRT/PMMA/Pt-TMD/SiO$_2$/Si stack is then placed into a water bath for 5-10 min for soaking. Then, the TRT/PMMA/Pt-TMD is carefully mechanically delaminated from the surface of SiO$_2$/Si. The process works best in water due to the specific interface of Pt-TMD and SiO$_2$: the TMDs are hydrophobic, while the SiO$_2$ is hydrophilic; therefore, upon peeling off, the water propagates into the bulk of the film, coats SiO$_2$/Si, and pushes the TMD out, assisting in the separation. When the TRT/PMMA/Pt-TMD is fully delaminated, it is dried slowly at room temperature. Before transferring Pt-TMD onto tattoo paper, the paper itself must be prepared. The tattoo paper comes with a thin layer of PVA on top, which must be removed. It is done by simply soaking the paper in water for 30 sec and picking up the partially dissolved PVA with tweezers. When the PVA is removed, the tattoo paper can be dried out by quick N$_2$ gun assisted drying and 10 min dry under ambient conditions. The TRT/PMMA/Pt-TMD is then brought in contact with the tattoo paper and placed on a hotplate at room temperature. The hotplate is then slowly heated-up to 120°C when the TRT is delaminated by itself, leaving the PMMA/Pt-TMD on top of the tattoo paper. In order to flip the PMMA/Pt-TMD to result in Pt-TMD facing up, it is simply flipped onto another tattoo paper. The material is then shaped into an arbitrary pattern with resolution down to 500 µm via mechanical plotter, Silhouette Cameo.[50] To transfer onto skin, the tattoo is soaked in water again, causing the paper to be slippery, allowing the transfer, and brought in contact with human skin. Sliding away the tattoo paper, Pt-TMD tattoo will stay on the skin. After the measurements, the tattoos can be very easily removed by gentle picking up via kind removal silicone tape for disposal.

**Fabrication of Soft Tapes and Ecoflex Slides**. In order to yield the thinnest possible electrical contacts that at the same time adhere to the skin, we utilized a 10 µm thick adhesive tape (Iwatani, ISR-BSMK10G) and evaporated a metal stack (Ni/Au, 10/90 nm) on top of it. The tape is supported by the PET backing layer, which is delaminated from PET right before placing the tape onto the skin.

To make the skin-resembling surface for resistance measurements, we coated glass slides with a thick (2-4 mm) layer of Ecoflex 00-30, which was cured at room temperature for 48 hours.

**Experiments on Human Subjects.** All experiments were conducted under the Institutional Review Board's approval at the University of Texas at Austin (protocol number: 2018-06-0058).

ACKNOWLEDGMENT

The work was supported in part by the Office of Naval Research grant #N00014-18-1-2706, the Temple Foundation Endowed Professorship, the National Science Foundation (CMMI-1728390), and the Creative Materials Discovery Program through the National Research Foundation of Korea (NRF) (NRF-2019M3D1A1069793).

REFERENCES

(1)    Zhou, J.; Lin, J.; Huang, X.; Zhou, Y.; Chen, Y.; Xia, J.; Wang, H.; Xie, Y.; Yu, H.; Lei, J.; Wu, D.; Liu, F.; Fu, Q.; Zeng, Q.; Hsu, C. H.; Yang, C.; Lu, L.; Yu, T.; Shen, Z.; Lin, H.; Yakobson, B. I.; Liu, Q.; Suenaga, K.; Liu, G.; Liu, Z. A Library of Atomically Thin Metal Chalcogenides. *Nature* **2018**, *556* (7701), 355–359. https://doi.org/10.1038/s41586-018-0008-3.

(2)    Manzeli, S.; Ovchinnikov, D.; Pasquier, D.; Yazyev, O. V.; Kis, A. 2D Transition Metal Dichalcogenides. *Nat. Rev. Mater.* **2017**, *2*. https://doi.org/10.1038/natrevmats.2017.33.

(3)    Kang, J.; Zhang, L.; Wei, S. H. A Unified Understanding of the Thickness-Dependent Bandgap Transition in Hexagonal Two-Dimensional Semiconductors. *J. Phys. Chem. Lett.* **2016**, *7* (4), 597–602. https://doi.org/10.1021/acs.jpclett.5b02687.

(4)    Villaos, R. A. B.; Crisostomo, C. P.; Huang, Z. Q.; Huang, S. M.; Padama, A. A. B.; Albao, M. A.; Lin, H.; Chuang, F. C. Thickness Dependent Electronic Properties of Pt Dichalcogenides. *npj 2D Mater. Appl.* **2019**, *3* (1), 1–8. https://doi.org/10.1038/s41699-018-0085-z.

(5)    Pang, Y.; Yang, Z.; Yang, Y.; Ren, T. L. Wearable Electronics Based on 2D Materials for Human Physiological Information Detection. *Small* **2020**, *16* (15), 1–26. https://doi.org/10.1002/smll.201901124.

(6)    Lee, J. Y.; Shin, J. H.; Lee, G. H.; Lee, C. H. Two-Dimensional Semiconductor Optoelectronics Based on van Der Waals Heterostructures. *Nanomaterials* **2016**, *6* (11), 40–43. https://doi.org/10.3390/nano6110193.

(7)    Pourbaix, M. J. N.; Van Muylder, J.; de Zoubov, N.; PROTECTION, C. Electrochemical Properties of the Platinum Metals. *Platin. Met. Rev.* **1959**, *3* (2), 47–53.

(8)    Greeley, J.; Stephens, I. E. L.; Bondarenko, A. S.; Johansson, T. P.; Hansen, H. A.; Jaramillo, T. F.; Rossmeisl, J.; Chorkendorff, I.; Nørskov, J. K. Alloys of Platinum and Early Transition Metals as Oxygen Reduction Electrocatalysts. *Nat. Chem.* **2009**, *1* (7), 552–556. https://doi.org/10.1038/nchem.367.

(9)    Geninatti, T.; Bruno, G.; Barile, B.; Hood, R. L.; Farina, M.; Schmulen, J.; Canavese, G.; Grattoni, A. Impedance Characterization, Degradation, and in Vitro Biocompatibility for Platinum Electrodes on BioMEMS. *Biomed. Microdevices* **2015**, *17* (1), 1–11. https://doi.org/10.1007/s10544-014-9909-6.

(10)   Zhao, Y.; Qiao, J.; Yu, Z.; Yu, P.; Xu, K.; Lau, S. P.; Zhou, W.; Liu, Z.; Wang, X.; Ji, W.; Chai, Y. High-Electron-Mobility and Air-Stable 2D Layered PtSe2 FETs. *Adv. Mater.* **2017**, *29* (5). https://doi.org/10.1002/adma.201604230.

(11)   Ciarrocchi, A.; Avsar, A.; Ovchinnikov, D.; Kis, A. Thickness-Modulated Metal-to-Semiconductor Transformation in a Transition Metal Dichalcogenide. *Nat. Commun.* **2018**, *9* (1), 1–6. https://doi.org/10.1038/s41467-018-03436-0.

(12) Yim, C.; Lee, K.; McEvoy, N.; O'Brien, M.; Riazimehr, S.; Berner, N. C.; Cullen, C. P.; Kotakoski, J.; Meyer, J. C.; Lemme, M. C.; Duesberg, G. S. High-Performance Hybrid Electronic Devices from Layered PtSe 2 Films Grown at Low Temperature. *ACS Nano* **2016**, *10* (10), 9550–9558. https://doi.org/10.1021/acsnano.6b04898.

(13) Okogbue, E.; Han, S. S.; Ko, T. J.; Chung, H. S.; Ma, J.; Shawkat, M. S.; Kim, J. H.; Kim, J. H.; Ji, E.; Oh, K. H.; Zhai, L.; Lee, G. H.; Jung, Y. Multifunctional Two-Dimensional PtSe2-Layer Kirigami Conductors with 2000% Stretchability and Metallic-to-Semiconducting Tunability. *Nano Lett.* **2019**, *19* (11), 7598–7607. https://doi.org/10.1021/acs.nanolett.9b01726.

(14) Han, S. S.; Kim, J. H.; Noh, C.; Kim, J. H.; Ji, E.; Kwon, J.; Yu, S. M.; Ko, T. J.; Okogbue, E.; Oh, K. H.; Chung, H. S.; Jung, Y.; Lee, G. H.; Jung, Y. Horizontal-to-Vertical Transition of 2D Layer Orientation in Low-Temperature Chemical Vapor Deposition-Grown PtSe 2 and Its Influences on Electrical Properties and Device Applications. *ACS Appl. Mater. Interfaces* **2019**, *11* (14), 13598–13607. https://doi.org/10.1021/acsami.9b01078.

(15) Guo, G. Y.; Liang, W. Y. The Electronic Structures of Platinum Dichalcogenides: PtS 2 , PtSe 2 and PtTe 2. *J. Phys. C Solid State Phys.* **1986**, *19* (7), 995–1008. https://doi.org/10.1088/0022-3719/19/7/011.

(16) Soled, S.; Wold, A.; Gorochov, O. Crystal Growth and Characterization of Platinum Ditelluride. *Mater. Res. Bull.* **1975**, *10* (8), 831–835. https://doi.org/10.1016/0025-5408(75)90199-3.

(17) Yan, M.; Huang, H.; Zhang, K.; Wang, E.; Yao, W.; Deng, K.; Wan, G.; Zhang, H.; Arita, M.; Yang, H.; Sun, Z.; Yao, H.; Wu, Y.; Fan, S.; Duan, W.; Zhou, S. Lorentz-Violating Type-II Dirac Fermions in Transition Metal Dichalcogenide PtTe2. *Nat. Commun.* **2017**, *8* (1), 1–6. https://doi.org/10.1038/s41467-017-00280-6.

(18) Okogbue, E.; Ko, T.-J.; Han, S. S.; Shawkat, M. S.; Wang, M.; Chung, H.-S.; Oh, K. H.; Jung, Y. Wafer-Scale 2D PtTe2 Layers for High-Efficiency Mechanically Flexible Electro-Thermal Smart Window Applications. *Nanoscale* **2020**. https://doi.org/10.1039/d0nr01845g.

(19) Wang, M.; Ko, T. J.; Shawkat, M. S.; Han, S. S.; Okogbue, E.; Chung, H. S.; Bae, T. S.; Sattar, S.; Gil, J.; Noh, C.; Oh, K. H.; Jung, Y.; Larsson, J. A.; Jung, Y. Wafer-Scale Growth of 2D PtTe2 with Layer Orientation Tunable High Electrical Conductivity and Superior Hydrophobicity. *ACS Appl. Mater. Interfaces* **2020**, *12* (9), 10839–10851. https://doi.org/10.1021/acsami.9b21838.

(20) Okogbue, E.; Han, S. S.; Ko, T.; Chung, H.; Ma, J.; Zhai, L.; Lee, G.; Jung, Y. Multifunctional Two-Dimensional PtSe2-Layer Kirigami Conductors with 2000% Stretchability and Metallic-to-Semiconducting Tunability. *Nano Lett.* **2019**, *19*, 7598–7607. https://doi.org/10.1021/acs.nanolett.9b01726.

(21) Kabiri Ameri, S.; Ho, R.; Jang, H.; Tao, L.; Wang, Y.; Wang, L.; Schnyer, D. M.; Akinwande, D.; Lu, N. Graphene Electronic Tattoo Sensors. *ACS Nano* **2017**, *11* (8), 7634–7641. https://doi.org/10.1021/acsnano.7b02182.

(22) Politano, A.; Chiarello, G.; Kuo, C.-N.; Lue, C. S.; Edla, R.; Torelli, P.; Pellegrini, V.; Boukhvalov, D. W. Tailoring the Surface Chemical Reactivity of Transition-Metal Dichalcogenide PtTe 2 Crystals. *Adv. Funct. Mater.* **2018**, *28* (15), 1706504. https://doi.org/10.1002/adfm.201706504.

(23) Sajjad, M.; Singh, N.; Schwingenschlögl, U. Strongly Bound Excitons in Monolayer PtS 2 and PtSe 2. *Appl. Phys. Lett.* **2018**, *112* (4), 043101. https://doi.org/10.1063/1.5010881.

(24) Hu, D.; Mendes, R. G.; Rümmeli, M. H.; Dai, Q.; Wu, B.; Fu, L.; Liu, Y. Highly Organized Epitaxy of Dirac Semimetallic PtTe2 Crystals with Extrahigh Conductivity and Visible Surface Plasmons at Edges. *ACS Nano* **2018**, *12* (9), 9405–9411. https://doi.org/10.1021/acsnano.8b04540.

(25) Hao, S.; Zeng, J.; Xu, T.; Cong, X.; Wang, C.; Wu, C.; Wang, Y.; Liu, X.; Cao, T.; Su, G.; Jia, L.; Wu, Z.; Lin, Q.; Zhang, L.; Yan, S.; Guo, M.; Wang, Z.; Tan, P.; Sun, L.; Ni, Z. Low-Temperature Eutectic Synthesis of PtTe 2 with Weak Antilocalization and Controlled Layer Thinning. **2018**, *1803746*, 1–9. https://doi.org/10.1002/adfm.201803746.

(26) Shi, J.; Huan, Y.; Hong, M.; Xu, R.; Yang, P.; Zhang, Z.; Zou, X.; Zhang, Y. Chemical Vapor Deposition Grown Large-Scale Atomically Thin Platinum Diselenide with Semimetal-Semiconductor Transition. *ACS Nano* **2019**, *13* (7), 8442–8451. https://doi.org/10.1021/acsnano.9b04312.

(27) Ko, T.-J.; Han, S. S.; Okogbue, E.; Shawkat, M. S.; Wang, M.; Ma, J.; Bae, T.-S.; Hafiz, S. Bin; Ko, D.-K.; Chung, H.-S.; Oh, K. H.; Jung, Y. Wafer-Scale 2D PtTe2 Layers-Enabled Kirigami Heaters with Superior Mechanical Stretchability and Electro-Thermal Responsiveness. *Appl. Mater. Today* **2020**, *20*, 100718. https://doi.org/10.1016/j.apmt.2020.100718.

(28) Cho, M. C.; Kim, J. Y.; Cho, S. H. A Bio-Impedance Measurement System for Portable Monitoring of Heart Rate and Pulse Wave Velocity Using Small Body Area. *Proc. - IEEE Int. Symp. Circuits Syst.* **2009**, *1*, 3106–3109. https://doi.org/10.1109/ISCAS.2009.5118460.

(29) Li, G.; Wang, S.; Duan, Y. Y. Towards Gel-Free Electrodes: A Systematic Study of Electrode-Skin Impedance. *Sensors Actuators, B Chem.* **2017**, *241*, 1244–1255. https://doi.org/10.1016/j.snb.2016.10.005.

(30) Ferrari, L. M.; Sudha, S.; Tarantino, S.; Esposti, R.; Bolzoni, F.; Cavallari, P.; Cipriani, C.; Mattoli, V.; Greco, F. Ultraconformable Temporary Tattoo Electrodes for Electrophysiology. *Adv. Sci.* **2018**, *5* (3), 1700771. https://doi.org/10.1002/advs.201700771.

(31) Gandhi, N.; Khe, C.; Chung, D.; Chi, Y. M.; Cauwenberghs, G. Properties of Dry and Non-Contact Electrodes for Wearable Physiological Sensors. *Proc. - 2011 Int. Conf. Body Sens. Networks, BSN 2011* **2011**, 107–112. https://doi.org/10.1109/BSN.2011.39.

(32) Beasley, C. Basics of Electrochemical Impedance Spectroscopy. *Webinar EIS Batter.* **2015**, No. 1. https://doi.org/10.1152/ajpregu.00432.2003.

(33)  Jeong, J. W.; Yeo, W. H.; Akhtar, A.; Norton, J. J. S.; Kwack, Y. J.; Li, S.; Jung, S. Y.; Su, Y.; Lee, W.; Xia, J.; Cheng, H.; Huang, Y.; Choi, W. S.; Bretl, T.; Rogers, J. A. Materials and Optimized Designs for Human-Machine Interfaces via Epidermal Electronics. *Adv. Mater.* **2013**, *25* (47), 6839–6846. https://doi.org/10.1002/adma.201301921.

(34)  Sel, K.; Kireev, D.; Brown, A.; Ibrahim, B.; Akinwande, D.; Jafari, R. Electrical Characterization of Graphene-Based e-Tattoos for Bio-Impedance-Based Physiological Sensing. *BioCAS 2019 - Biomed. Circuits Syst. Conf. Proc.* **2019**, 1–4. https://doi.org/10.1109/BIOCAS.2019.8919003.

(35)  Kireev, D.; Offenhaeusser, A.; Offenhäusser, A. Graphene & Two-Dimensional Devices for Bioelectronics and Neuroprosthetics. *2D Mater.* **2018**, *5* (4), 042004. https://doi.org/10.1088/2053-1583/aad988.

(36)  Bjerregaard, P.; Gussak, I. Naming of the Waves in the ECG With a Brief Account of Their Genesis. *Circulation* **1999**, *100* (25), 1937–1942. https://doi.org/10.1161/01.CIR.100.25.e148.

(37)  Sadikoglu, F.; Kavalcioglu, C.; Dagman, B. Electromyogram (EMG) Signal Detection, Classification of EMG Signals and Diagnosis of Neuropathy Muscle Disease. *Procedia Comput. Sci.* **2017**, *120*, 422–429. https://doi.org/10.1016/j.procs.2017.11.259.

(38)  Vaid, S.; Singh, P.; Kaur, C. EEG Signal Analysis for BCI Interface: A Review. *Int. Conf. Adv. Comput. Commun. Technol. ACCT* **2015**, *2015-April*, 143–147. https://doi.org/10.1109/ACCT.2015.72.

(39)  Vidal, F.; Burle, B.; Spieser, L.; Carbonnell, L.; Meckler, C.; Casini, L.; Hasbroucq, T. Linking EEG Signals, Brain Functions and Mental Operations: Advantages of the Laplacian Transformation. *Int. J. Psychophysiol.* **2015**, *97* (3), 221–232. https://doi.org/10.1016/j.ijpsycho.2015.04.022.

(40)  Oostenveld, R.; Praamstra, P. The Five Percent Electrode System for High-Resolution EEG and ERP Measurements. *Clin. Neurophysiol.* **2001**, *112* (4), 713–719. https://doi.org/10.1016/S1388-2457(00)00527-7.

(41)  Li, L. The Differences among Eyes-Closed, Eyes-Open and Attention States: An EEG Study. *2010 6th Int. Conf. Wirel. Commun. Netw. Mob. Comput. WiCOM 2010* **2010**, No. Nsfc 30800242. https://doi.org/10.1109/WICOM.2010.5600726.

(42)  Bulling, A.; Ward, J. A.; Gellersen, H.; Tröster, G. Eye Movement Analysis for Activity Recognition Using Electrooculography. *IEEE Trans. Pattern Anal. Mach. Intell.* **2011**, *33* (4), 741–753. https://doi.org/10.1109/TPAMI.2010.86.

(43)  Usakli, A. B.; Gurkan, S.; Aloise, F.; Vecchiato, G.; Babiloni, F. On the Use of Electrooculogram for Efficient Human Computer Interfaces. *Comput. Intell. Neurosci.* **2010**, *2010*, 1–5. https://doi.org/10.1155/2010/135629.

(44)  Kim, D. Y.; Han, C. H.; Im, C. H. Development of an Electrooculogram-Based Human-Computer Interface Using Involuntary Eye Movement by Spatially Rotating Sound for Communication of Locked-in Patients. *Sci. Rep.* **2018**, *8* (1), 1–10. https://doi.org/10.1038/s41598-018-27865-5.

(45)　Barea Navarro, R.; Boquete Vázquez, L.; López Guillén, E. EOG-Based Wheelchair Control. In *Smart Wheelchairs and Brain-Computer Interfaces*; Elsevier, 2018; pp 381–403. https://doi.org/10.1016/B978-0-12-812892-3.00016-9.

(46)　Gao, J. F.; Yang, Y.; Lin, P.; Wang, P.; Zheng, C. X. Automatic Removal of Eye-Movement and Blink Artifacts from Eeg Signals. *Brain Topogr.* **2010**, *23* (1), 105–114. https://doi.org/10.1007/s10548-009-0131-4.

(47)　Open BCI. Ganglion Getting Started Guide https://docs.openbci.com/docs/01GettingStarted/01-Boards/GanglionGS.

(48)　Ibrahim, B.; McMurray, J.; Jafari, A. R. A Wrist-Worn Strap with an Array of Electrodes for Robust Physiological Sensing. *Proc. Annu. Int. Conf. IEEE Eng. Med. Biol. Soc. EMBS* **2018**, *2018-July*, 4313–4317. https://doi.org/10.1109/EMBC.2018.8513238.

(49)　Zahniser, D. J.; Brenner, J. F. Signals and Systems, by A.V. Oppenheim, A.S. Willsky, and I.T. Young. Prentice-Hall, Englewood Cliffs, New Jersey, 1983, 796 Pages, Hardbound, $37.95. *Cytometry* **1985**, *6* (4), 392–392. https://doi.org/10.1002/cyto.990060420.

(50)　Yang, S.; Chen, Y.-C.; Nicolini, L.; Pasupathy, P.; Sacks, J.; Su, B.; Yang, R.; Sanchez, D.; Chang, Y.-F.; Wang, P.; Schnyer, D.; Neikirk, D.; Lu, N. "Cut-and-Paste" Manufacture of Multiparametric Epidermal Sensor Systems. *Adv. Mater.* **2015**, *27* (41), 6423–6430. https://doi.org/10.1002/adma.201502386.

# Supporting Information

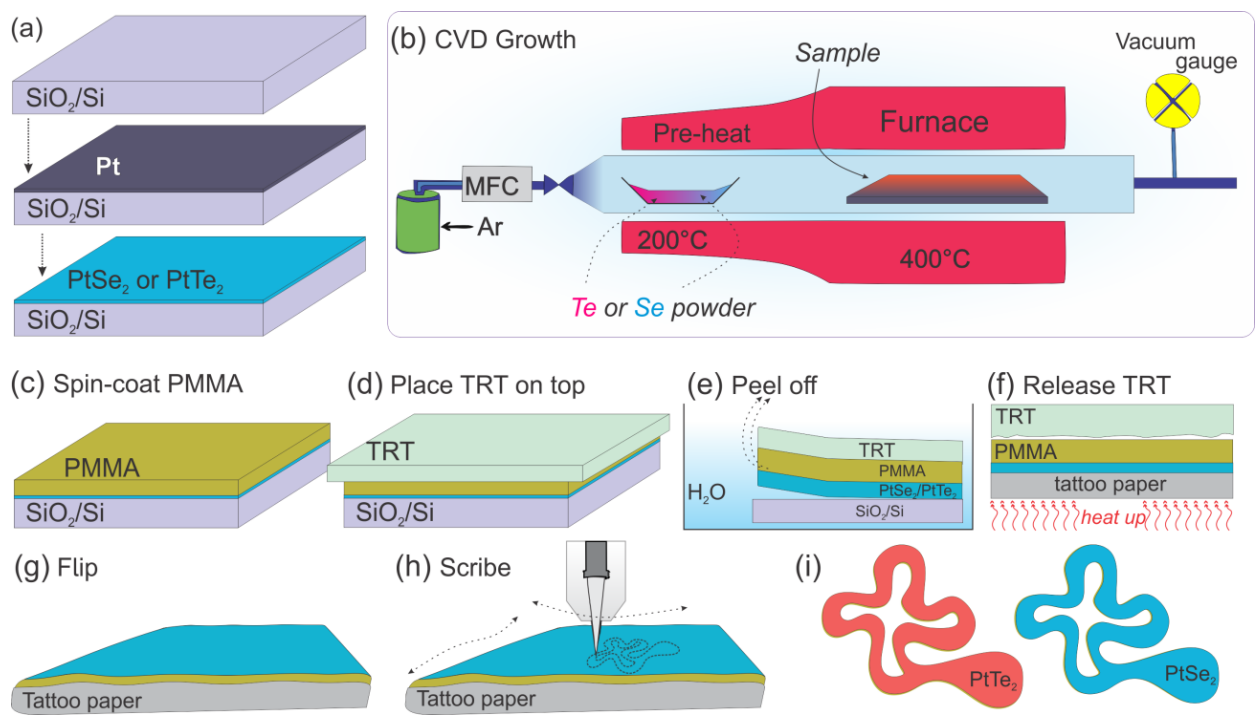

Figure S1. Schematic overview of the PMMA-supported ultrathin Pt-TMD tattoo design & fabrication flow. (a) Shows the evaporation of thin Pt on top of the $SiO_2/Si$ wafer, followed by TAC conversion into Pt-TMD. (b) Shows schematic of the TAC CVD process. (c) Post CVD growth, a thin layer of PMMA is spin-coated on top of the Pt-TMD/$SiO_2$/Si sample, followed by (d) attaching TRT on the top of the stack for mechanical support. (e) The stack is then soaked in water and the TRT is used to peel off the material from the surface of the $SiO_2$/Si. The process is 100% successful when performed in water due to pre-treatment of the $SiO_2$/Si wafer with oxygen plasma prior to the Pt evaporation and TAC process. The oxygen plasma makes the $SiO_2$ surface to become highly hydrophilic and since $PtSe_2$ and $PtTe_2$ materials are hydrophobic, the separation works faster and fault-free. (f-g) Releasing PMMA/Pt-TMD from TRT by placing them over a tattoo paper and heating up, followed by a flipping procedure to yield Pt-TMD/PMMA structure. (h) The scribing process, in the case of Pt-TMD/PMMA/tattoo paper. (i) Schematics of the final PtSe$_2$ and PtTe$_2$ tattoos supported by PMMA.

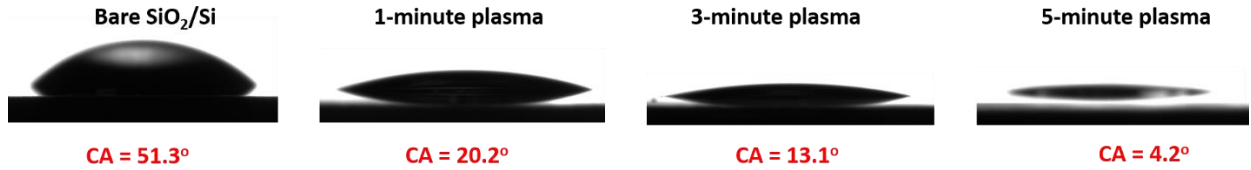

Figure S2. Water contact angle of SiO$_2$ surface before and after oxygen plasma treatment of 1-, 3-, and 5- minutes duration.

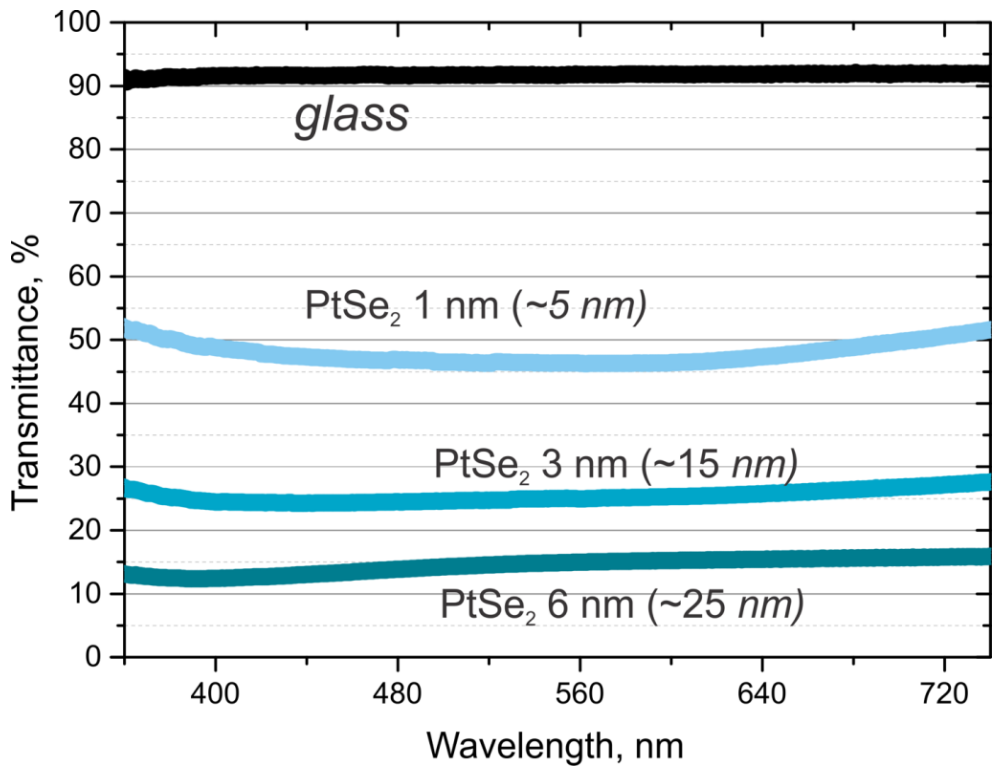

Figure S3. Visible light transmittance of the PtSe$_2$ films of varied thickness, used in this work.

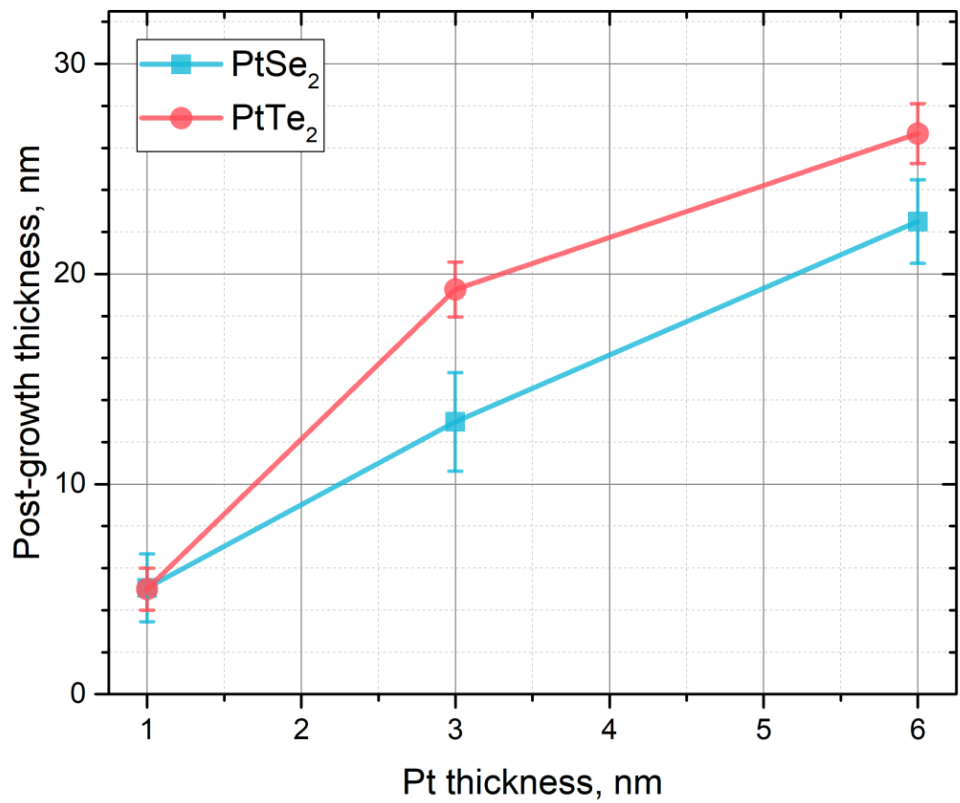

Figure S4. The correlation between as-deposited Pt layer (x-axis) and final post CVD-growth

2D material thickness, as per AFM measurements.

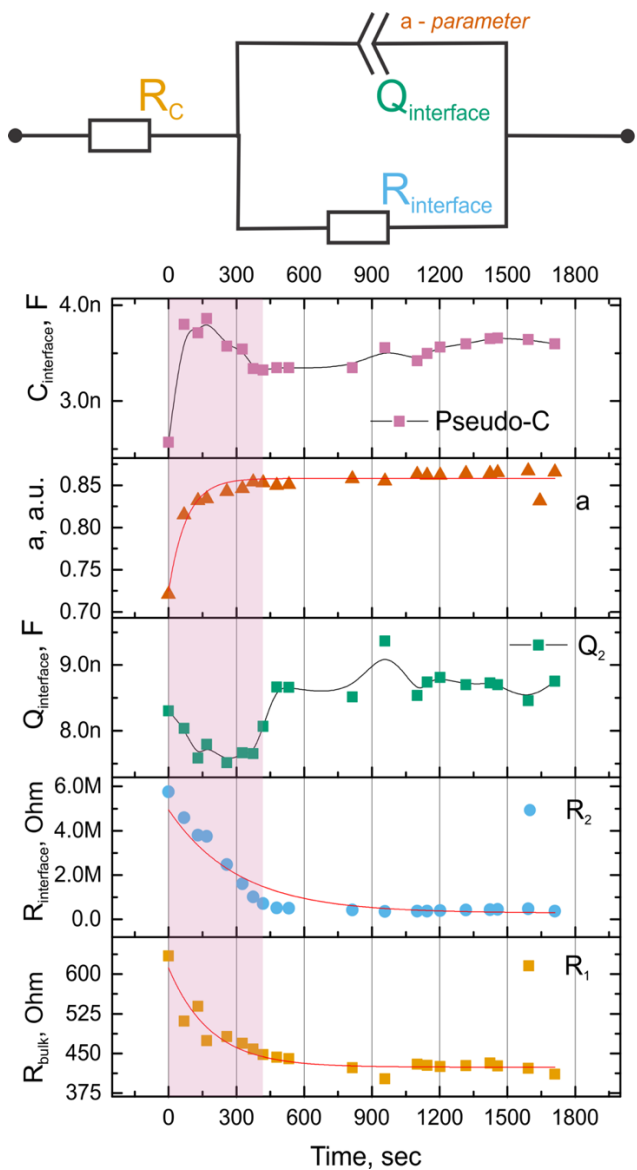

Figure S5. The simplified equivalent circuit (top), and variation of the fitted values with time (bottom). $R_{bulk}$ and $R_{interface}$ both asymptotically reduce their values with time, down to almost 425 Ohm and 285 kOhm accordingly. The interface capacitance in the model is represented by a constant phase element rather than a perfect capacitor, therefore there are 3 values that we report: a-the "perfectness" of the capacitor, Q – the CPE capacitance, and C, or rather pseudo capacitance of the interface that is calculated by taking care the value of a into consideration.

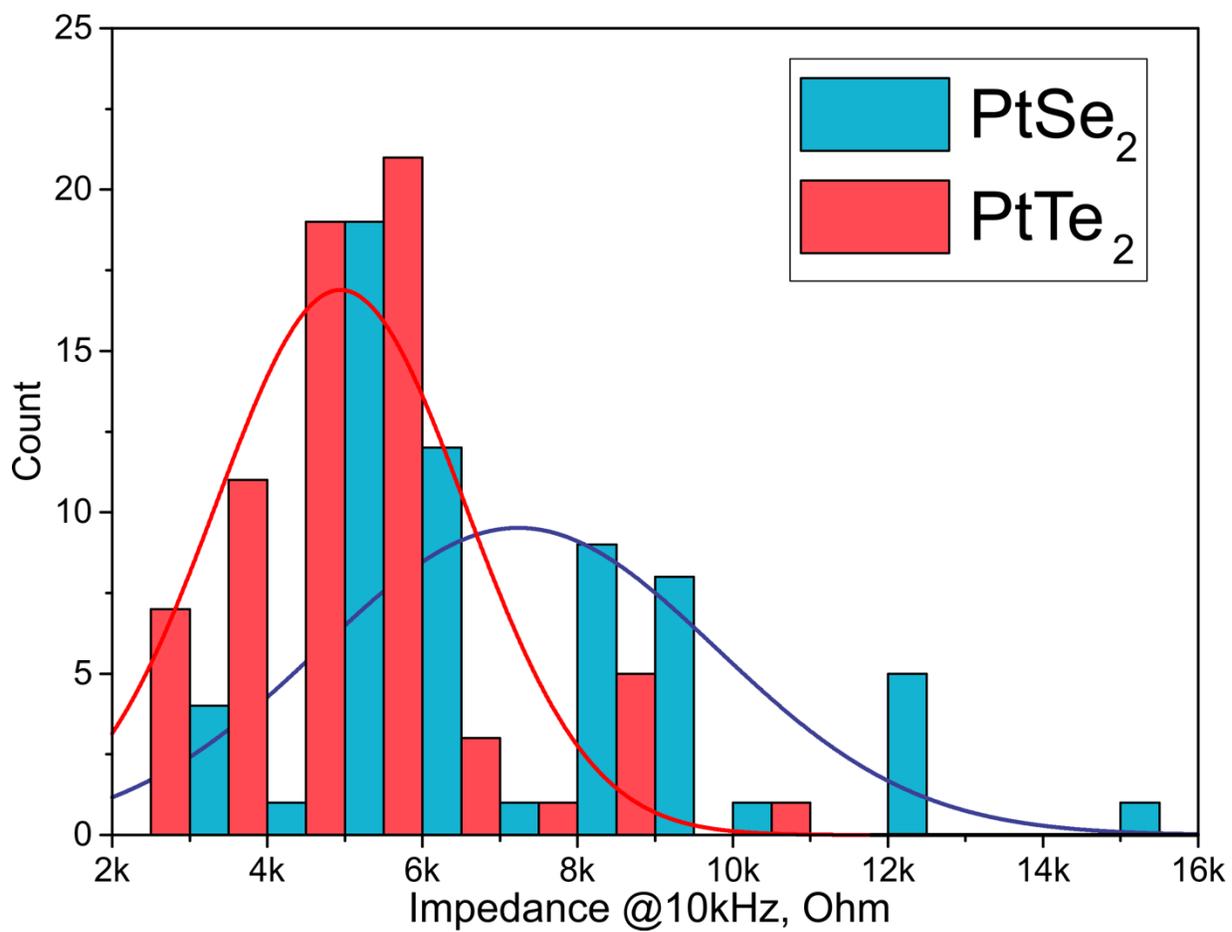

Figure S6. The statistical distribution of the electrode-skin impedance values (@10 kHz) for PtSe$_2$ and PtTe$_2$ electronic tattoos.

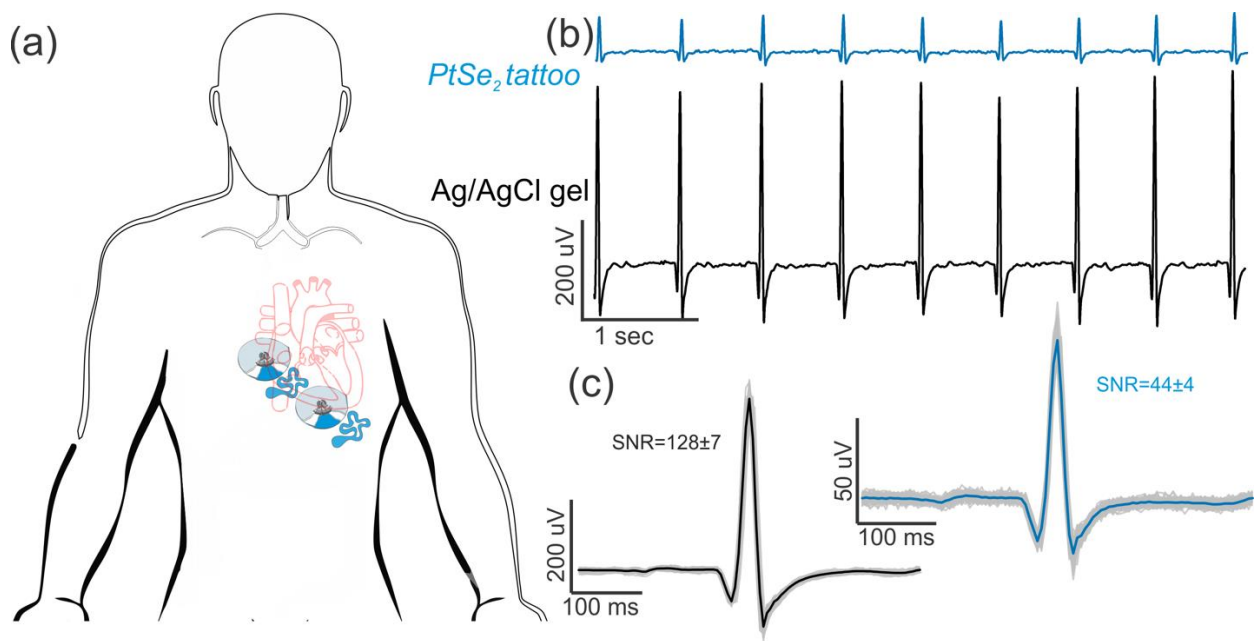

Figure S7. ECG recordings with PtSe$_2$ tattoos (blue) and comparison to the pair of Ag/AgCl gel electrodes (black). While the signal amplitude of Ag/AgCl is much higher, the noise is also larger. The calculated SNR, even for the PtSe$_2$ case, is acceptably large.

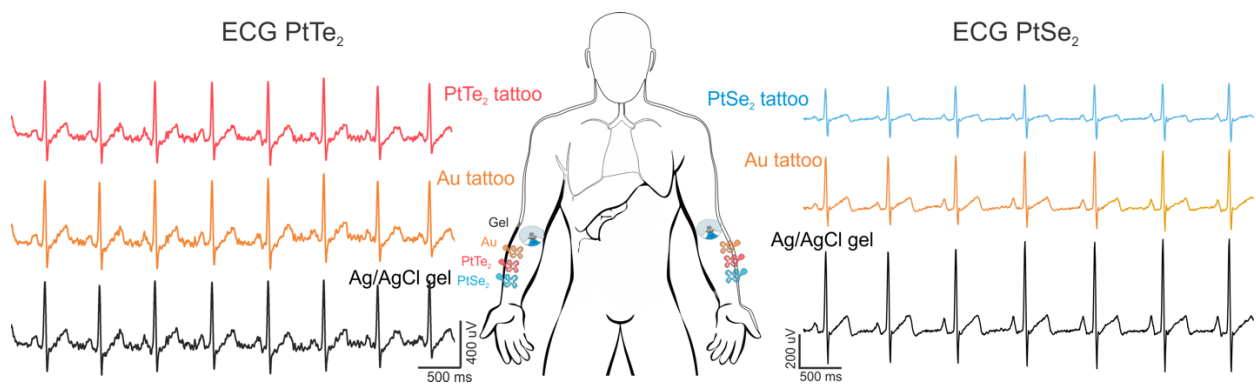

Figure S8. ECG recordings with PtSe$_2$ and PtTe$_2$ as recorded from two hands instead of a chest. Due to large spacing, the PQRST peaks are clearly distinguishable for all, PtSe$_2$, PtTe$_2$, gold, and Ag/AgCl gel electrodes.

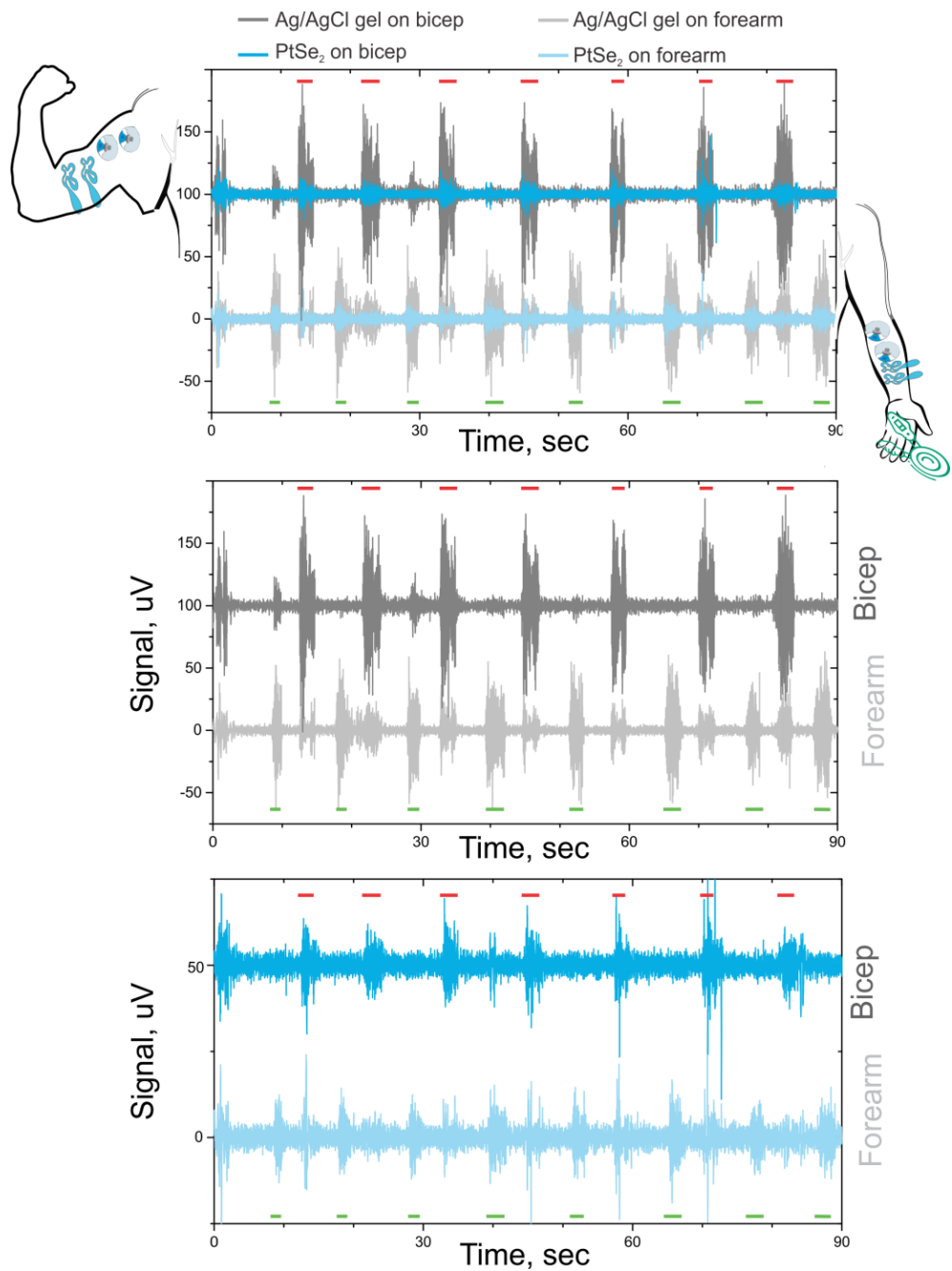

Figure S9. Electromyogram recordings from a human's forearm (bottom, light gray and light blue), and bicep (dark gray and dark blue) muscle contractions recorded via the Ag/Ag gel electrodes and **PtSe₂** tattoos correspondingly. The horizontal red bars on the top and green bars at the bottom mark the events and durations of the bicep curls and forearm contractions correspondingly.

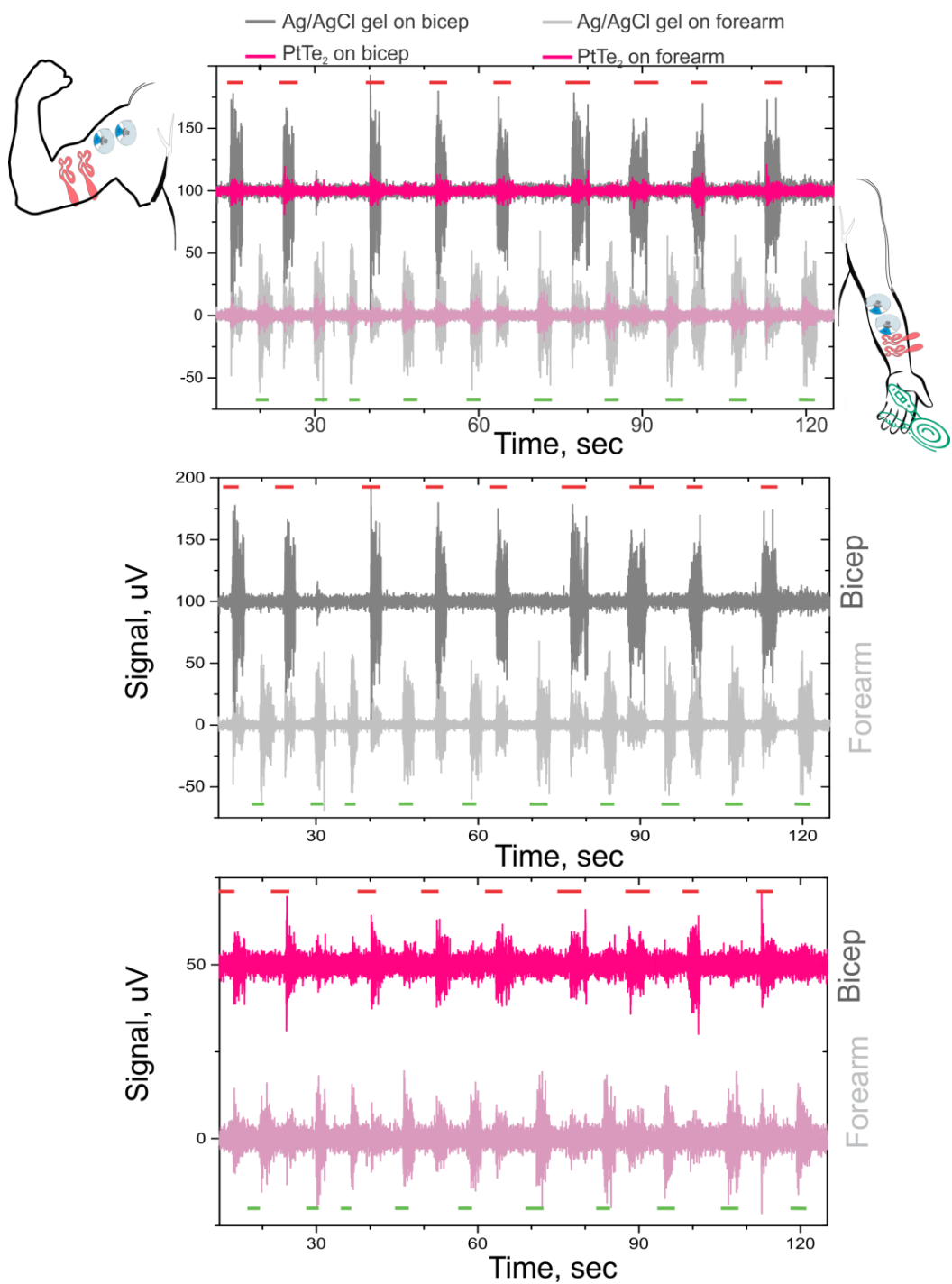

Figure S10. EMG recordings from a human's forearm (bottom, light gray and light pink), and bicep (dark gray and dark pink) muscle contractions recorded via the Ag/Ag gel electrodes and PtTe₂ tattoos correspondingly. The horizontal red bars on the top and green bars at the bottom mark the events and durations of the bicep curls and forearm contractions correspondingly.

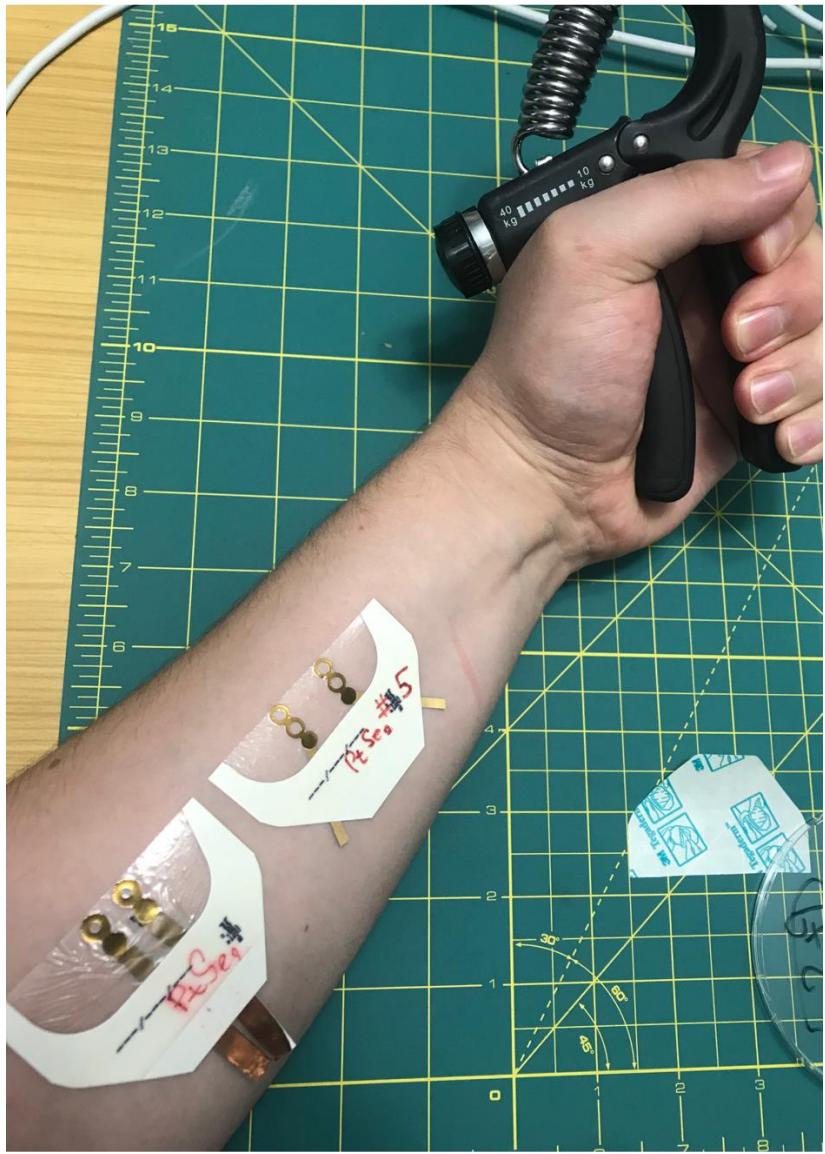

Figure S11. Photograph of two EMG recording tattoo pairs, both with PtSe₂, temporary supported with Tegaderm for stability and reusability.

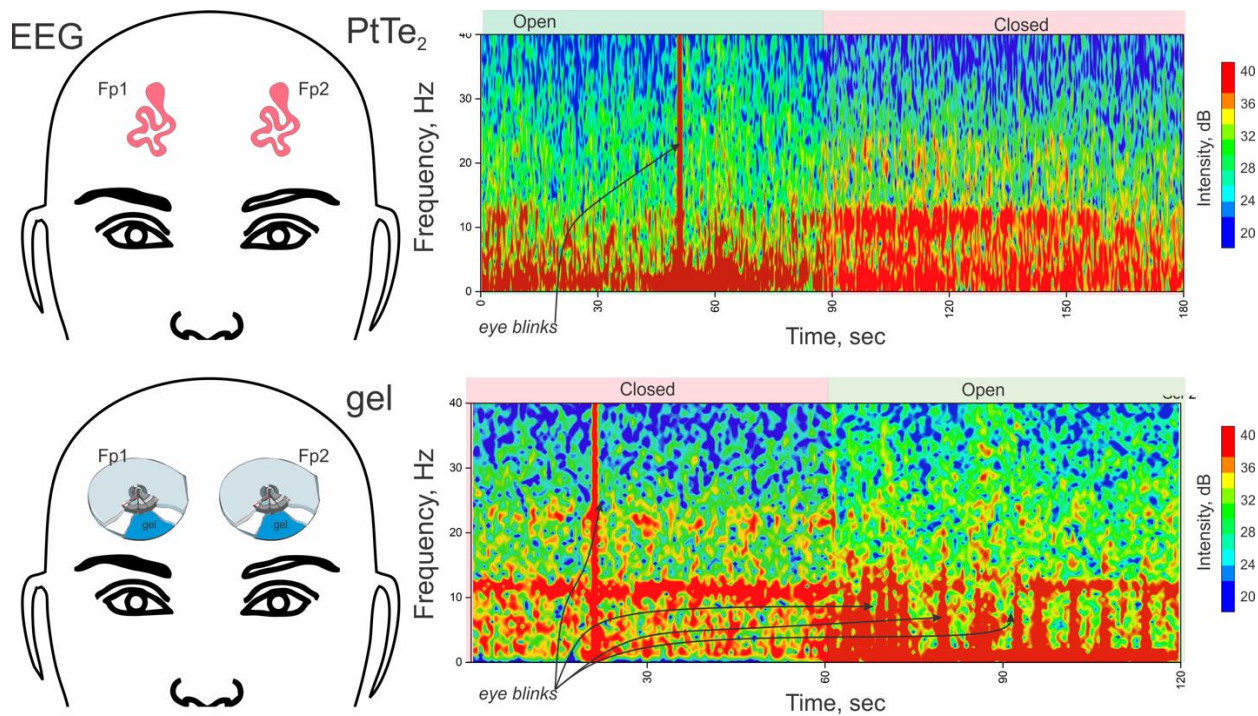

Figure S12. The electrode placement for EEG measurements (left) and resulting spectrogram (right) from the PtTe$_2$ tattoos (top) and Ag/AgCl gel electrodes (bottom) that are placed on the Fp1 and Fp2 locations on the subject's forehead. The eye closed-open pattern is clearly distinguishable by the presence-absence of alpha waves (8-13 Hz).

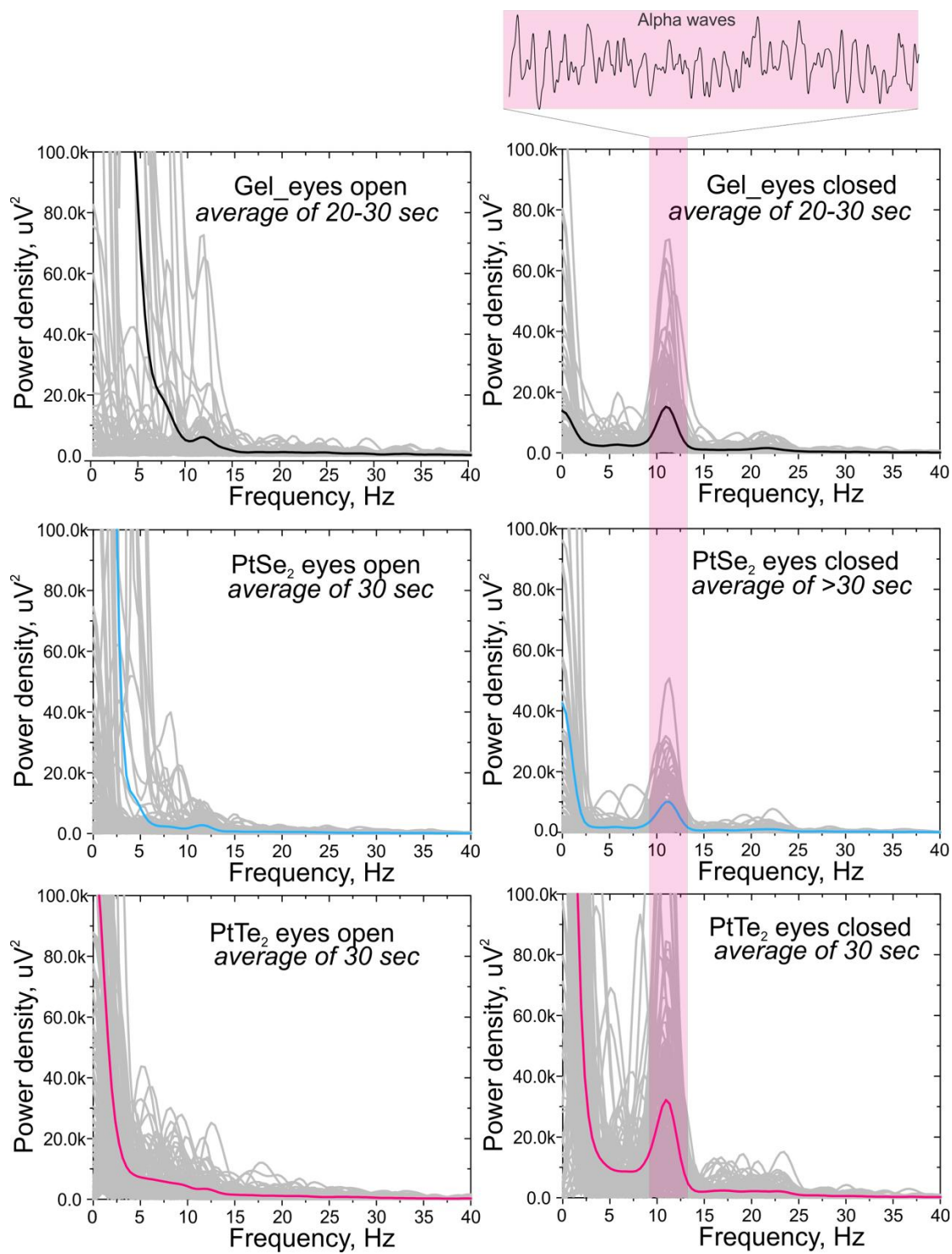

Figure S13. Power spectral density from gel, PtSe₂, and PtTe₂ electrodes with open (left) and closed (right) eyes. As seen from the FFT-ed data, with the eyes closed there is a significant power density peak at around 8-13 Hz, which corresponds to the alpha waves.

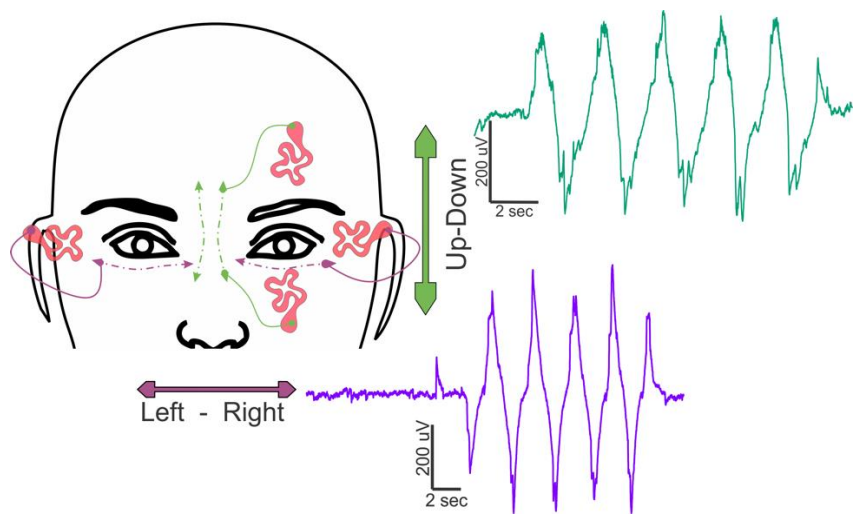

Figure S14. EOG tattoo placement, and associated signals recorded from the specific channels, green channel for rapid up-down eye movements, and violet for left-right eye movements.

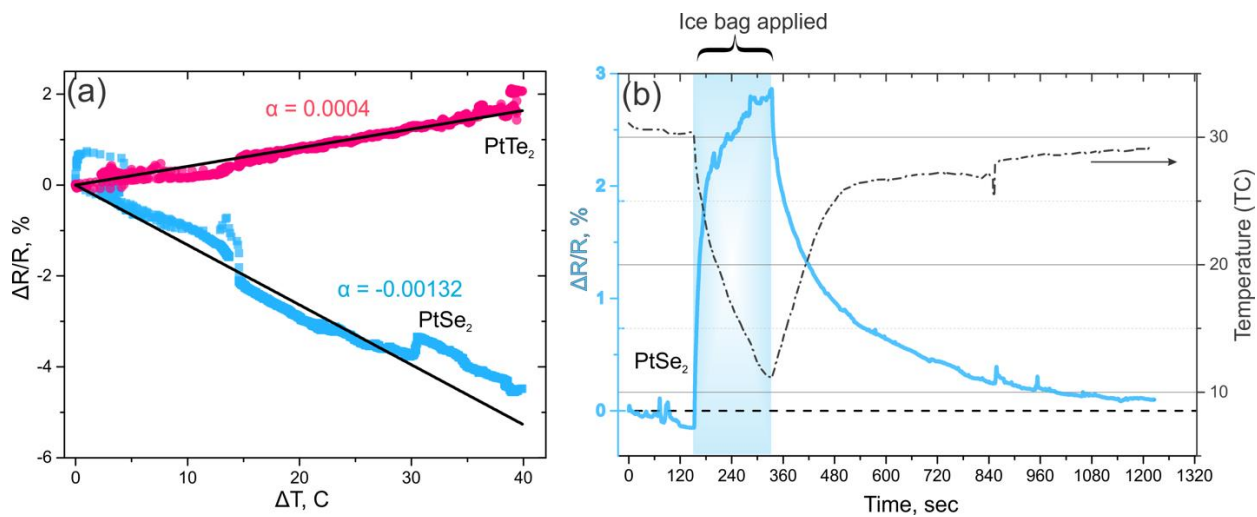

Figure S15. (a) Typical TEC responses of PtTe$_2$ (pink) and PtSe$_2$ (blue) tattoos. The fitting curves (black) are plotted with R$^2$ value above 98%. (b) Timetrace resistance response upon application of ice onto the body, as measured by the PtSe$_2$ tattoo (blue) and a thermocouple (black).

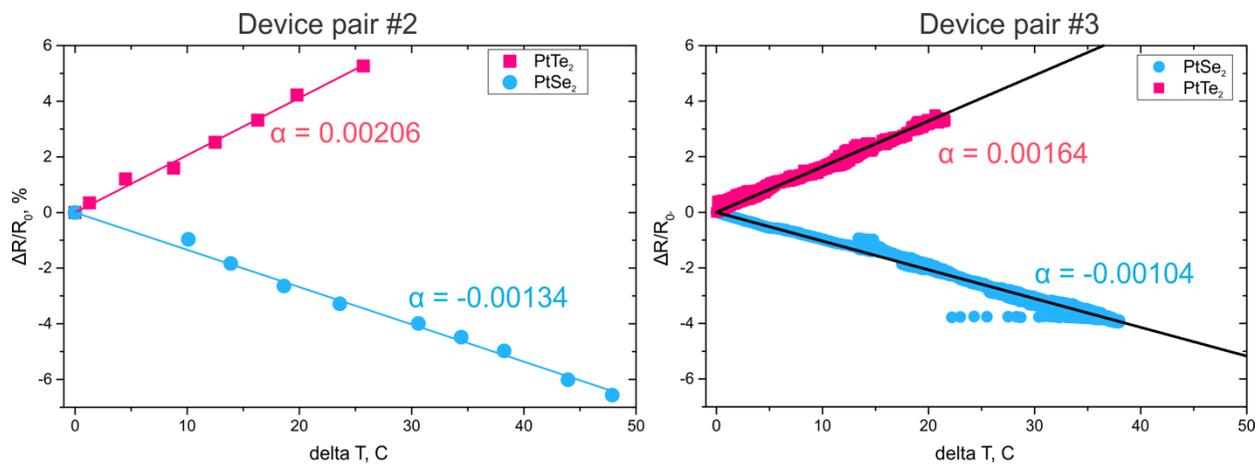

Figure S16. Two other batches of PtSe$_2$ and PtTe$_2$ used for temperature coefficient test. Similar to the other samples, PtTe$_2$ features positive TEC, while PtSe$_2$ shows negative sign of TEC.

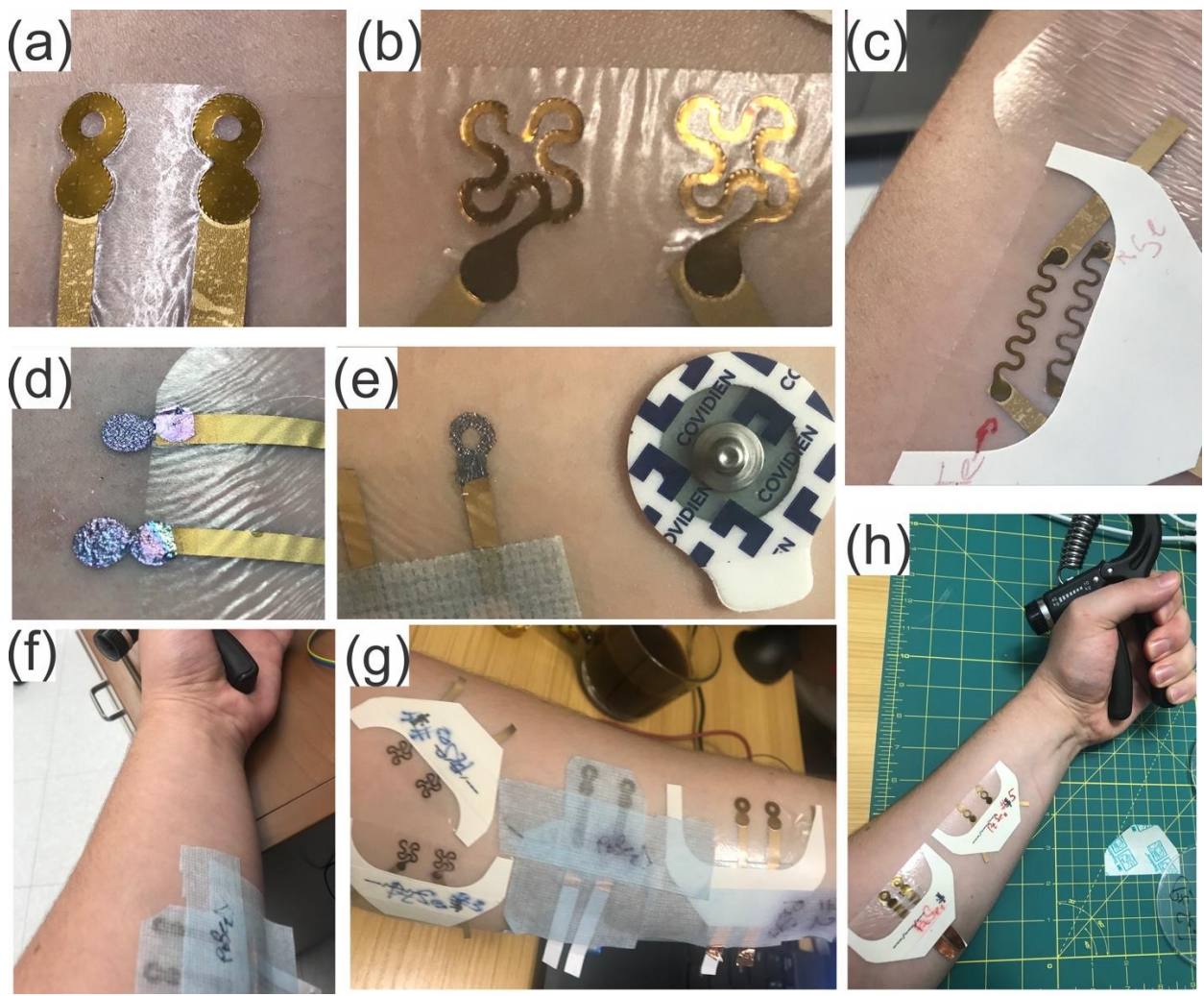

Figure S17. Optical images of the different kinds of PtSe$_2$ and PtTe$_2$ tattoos, either supported by Kapton (a-c, f-h) or PMMA (d-e). The (a) and (b) are pairs of the tattoos placed on a forearm used for skin impedance measurements. (c) shows two serpentine shaped stripes of PtSe$_2$ and PtTe$_2$ as used for temperature monitoring. (d-e) shows size comparison of PMMA-supported Pt-TMD tattoos to the classical Ag/AgCl gel electrodes. (f-h) are different combinations of multiple tattoos placed on a forearm for EMG monitoring.